\documentclass[prx,amsfonts,floatfix,superscriptaddress,longbibliography,twocolumn]{revtex4-1}
\usepackage{mathbbol}
\usepackage{graphicx,color}
\usepackage{amsmath,amssymb,bm,amsthm}
\usepackage{epstopdf}

\usepackage{bbding}

\def\be{\begin{equation}}
\def\ee{\end{equation}}
\def\ba{\begin{eqnarray}}
\def\ea{\end{eqnarray}}

\begin{document}
\title{Self-averaging in many-body quantum systems out of equilibrium. II. \\Approach to the localized phase}

\author{E. Jonathan Torres-Herrera}
\address{Instituto de F\'isica, Benem\'erita Universidad Aut\'onoma de Puebla,
Apt. Postal J-48, Puebla, 72570, Mexico}
\author{Giuseppe De Tomasi}
\address{T.C.M. Group, Cavendish Laboratory, JJ Thomson Avenue, Cambridge CB3 0HE, United Kingdom}
\author{Mauro Schiulaz}
\address{Department of Physics, Yeshiva University, New York City, New York, 10016, USA}
\author{Francisco P\'erez-Bernal}
\address{Dep. CC. Integradas y Centro de Estudios Avanzados en F\'isica,
Matem\'aticas y Computaci\'on. Fac. CC. Experimentales, Universidad de Huelva, Huelva 21071, \& Instituto Carlos I de F\'isica Te\'orica y Computacional, Universidad de Granada, Granada 18071, Spain}
\author{Lea F. Santos}
\address{Department of Physics, Yeshiva University, New York City, New York, 10016, USA}

\date{\today}

\begin{abstract}
The self-averaging behavior of interacting many-body quantum systems has been mostly studied at equilibrium. The present work addresses what happens out of equilibrium, as the increase of the strength of onsite disorder takes the system to the localized phase. We consider two local and two non-local quantities of great experimental and theoretical interest. In the delocalized phase, self-averaging depends on the observable and on the time scale, but the picture simplifies substantially when localization is reached. In the localized phase, the local observables become self-averaging at all times, while the non-local quantities are throughout non-self-averaging. These behaviors are explained and scaling analysis are provided using the $\ell$-bits model and a toy model.
\end{abstract}

\maketitle

\section{Introduction}

A central question in studies of disordered systems is whether self-averaging holds or not~\cite{LifshitzBook}. A quantity is self-averaging when the ratio between its variance over disorder realizations and the square of its average decreases with system size~\cite{Wiseman1995,Aharony1996,Wiseman1998,Castellani2005,Malakis2006,Roy2006,Monthus2006,Efrat2014,Lobejko2018}. This implies that the number of samples used in an experiment or statistical analyses can be reduced as the system size increases. It also means that in the thermodynamic limit, the quantity's behavior does not depend on the particular disorder realization used. If self-averaging does not hold, averages over large sets of random realizations are needed no matter how large the system is.

The analysis of self-averaging is commonly done in association with studies of normal and anomalous diffusion~\cite{Bouchaud1990,AkimotoPRL2016,Russian2017,AkimotoPRE2018} and transitions into the spin-glass state~\cite{Pastur1990,Wreszinski2004,Castellani2005}. Various quantities have been investigated, from susceptibility, specific heat, conductance, and free energy to entanglement entropy. At critical points, self-averaging is usually absent~\cite{Wiseman1995,Aharony1996,Wiseman1998, Parisi2002,Castellani2005,Malakis2006,Roy2006,Monthus2006,Efrat2014,MullerARXIV,Pastur2014PRL}. 

The present work focuses on the self-averaging properties of interacting many-body quantum systems with onsite disorder. In one-dimension, these systems exhibit two regimes, a delocalized phase when the disorder strength is smaller than the interaction strength and a localized phase reached when the disorder strength exceeds a critical point~\cite{SantosEscobar2004,Basko2006,Nandkishore2015,Luitz2017,Altman2018}. The phenomenon of many-body localization has been intensely examined, but several questions remain open, including the exact value of the critical point~\cite{Doggen2018}.

In the delocalized phase where the system is chaotic, the spectrum shows level statistics as in full random matrices. In this phase, one often assumes that self-averaging holds, so as the system size increases, one reduces the number of disorder realizations used in numerical simulations. More care is usually taken in the vicinity of the transition to the localized phase, for which several discussions exist about the lack of self-averaging~\cite{Barisic2016,Prelovsek2017,Khemani2017,Serbyn2017}. Most of these works address self-averaging behavior at equilibrium. The few papers that target self-averaging properties out of equilibrium include studies about spin-spin correlations~\cite{Richter2020}, reduced density matrices of embedded quantum systems~\cite{Ithier2017}, and driven systems~\cite{Lobejko2018,Mukherjee2018}. There are also studies about the lack of self-averaging of the two-level form factor~\cite{Argaman1993b,Eckhardt1995,Prange1997,Braun2015}, which is not an actual dynamical quantity, but an alternative to analyze spectral properties in the time domain

When we first approached the subject of self-averaging, our goal was to analyze what happens to the self-averaging behavior of interacting many-body quantum systems out of equilibrium. The plan was to investigate how the behavior of different quantities and at different disorder strengths might depend on the time scales. To our surprise, we found analytically and confirmed numerically that even in the chaotic regime, some quantities are non-self-averaging. Our first paper was then entirely dedicated to the chaotic regime~\cite{Schiulaz2020}, using for this a fixed value of the disorder strength.

We now come back to our original goal and investigate the self-averaging properties of an interacting spin model out of equilibrium as the disorder strength increases and the system approaches a many-body localized phase. We consider four quantities that have been extensively studied in nonequilibrium quantum dynamics: survival probability, inverse participation ratio, spin autocorrelation function, and connected spin-spin correlation function. The first two are non-local quantities in real space and the last two are local observables. The spin autocorrelation function is equivalent to the density imbalance used in experiments with cold atoms~\cite{Schreiber2015} and the connected spin-spin correlation function is measured in experiments with ion traps~\cite{Richerme2014}. 

In the chaotic regime, the results are non-trivial and highly dependent on the quantity and time scale~\cite{Schiulaz2020}. The same quantity may be non-self-averaging at short times, but self-averaging at long times, or vice versa. As the disorder strength increases and the system approaches localization, the crossing point for the opposing behaviors gets postponed to ever longer times. In the localized phase, the general picture becomes rather simple: the local quantities are self-averaging at any time scale and the non-local quantities are non-self-averaging at all times. 

We present numerical results and justifications for the changes that take place in the self-averaging behavior as the disorder strength grows. For the many-body localized phase, we use the $\ell$-bits model to show that the relative variances of the local observables are proportional to the reciprocal of the system size. This effective model represents a mapping of the interacting many-body system into an integrable system with an extensive number of integrals of motion~\cite{Serbyn2013,Huse2014,Iemini2016,Tomasi2019}. We also employ a toy model to explain why the relative variance of the global quantities grow exponentially with system size. Both models allow us to access system sizes much larger than those reachable with the exact diagonalization of the original spin Hamiltonian.

The paper is organized as follows. Self-averaging is defined in Sec.~\ref{Sec:Def}, where we present also the Hamiltonian, initial states, observables, $\ell$-bits model, and toy model. The next four sections, Secs.~III, IV, V, and VI, are then dedicated to the dependence of the relative variance of the four quantities considered on the disorder strength and on time. The first paragraph in each one of these sections summarizes the main findings. A reader interested only in the results for the many-body localized phase and the scaling analysis obtained with the $\ell$-bits model may skip directly to Sec.~\ref{Sec:lbit}. Conclusions are given in Sec.~\ref{Sec:Conc}.

\section{Self-averaging, model, and observables}
\label{Sec:Def}

This section defines self-averaging and the quantities and spin Hamiltonian studied. It also presents the models used to describe the localized phase, the $\ell$-bits and a toy model.

\subsection{Self-Averaging}

A quantity $O$ is self-averaging when its relative variance, which corresponds to the ratio between its variance $\sigma^2_{O}$ over disorder realizations and the square of its mean, that is,
\be
{\cal R}_{O}(t)=\frac{\sigma^2_{O}(t)}{\left<O(t)\right>^2} = \frac{\left<O^2(t)\right>-\left<O(t)\right>^2}{\left<O(t)\right>^2} ,
\label{eq:sigma} 
\ee
goes to zero as the system size $L$ increases. The notation $\langle.\rangle$ in the equation above indicates average over disorder realizations, and in our case, it includes also average over initial states. These states are taken in a very narrow window of energy around the center of the spectrum. The decrease of the relative variance with $L$ implies that in the thermodynamic limit, the sample to sample fluctuations vanish. 

Strong self-averaging refers to the case where ${\cal R}_{O}(t) \sim L^{-1}$, and weak self-averaging means that ${\cal R}_{O}(t) \sim L^{-\nu}$ with $0<\nu<1$. In many-body quantum systems, where the initial state can eventually spread over an exponentially large many-body Hilbert space, one can also encounter what we call ``super" self-averaging, when the relative variance decreases exponentially with the system size~\cite{Schiulaz2020}. 

The question addressed by the standard definition of self-averaging in Eq.~(\ref{eq:sigma}) is whether the variance of the quantity $O$ goes to zero faster than $\left<O(t)\right>^2$. The square of the mean value serves as a reference to determine whether the variance is large or small. If the mean value is exactly zero, independently of the system size, then the analysis of self-averaging should be done based on the value of the variance only, not the ratio. In our case, none of the quantities considered have mean zero, but they might approach zero at long times as the system size increases, so Eq.~(\ref{eq:sigma}) is the proper measure to use.

We emphasize that self-averaging is a concept intrinsically related with the presence of randomness in the Hamiltonian. The relative variance, ${\cal R}_{O}(t)$, that we study here involves averages over disorder realizations. This is different from relative variances involving temporal averages, 
\be
{\cal T}_{O}=\frac{\overline{O^2}-\overline{O}^2}{\overline{O}^2} ,
\label{Eq:temporal}
\ee
where
\be
\overline{O} = \lim_{T \rightarrow \infty} \frac{1}{T} \int_{0}^T  O(t) dt.
\label{Eq:infinite}
\ee
While ${\cal R}_{O}(t)$ depends on time, ${\cal T}_{O}$, is time independent and may be obtained for a single realization. ${\cal T}_{O}$ has been employed in studies of equilibration and thermalization~\cite{Srednicki1996,Reimann2008,Short2011,Zangara2013,TorresKollmar2015,Nation2019ARXIV}, and also many-body localization~\cite{Serbyn2014,Tomasi2019}.

Equilibration happens after the relaxation time $t_{\text{R}}$, when the dynamics finally saturates and the observable simply fluctuates around its infinite time average $\overline{O}$ \cite{Srednicki1996,Reimann2008,Short2011,Zangara2013,TorresKollmar2015,Zangara2013,Nation2019ARXIV,Dymarsky2019}. At this large time scales, $t>t_{\text{R}}$, one may expect ${\cal R}_{O}(t)$ at a fixed $t$ to coincide with ${\cal T}_{O}$ when the system is chaotic, due to ergodicity. Whether a relationship between ${\cal R}_{O}(t>t_{\text{R}})$ and  ${\cal T}_{O}$ might exist also for larger disorder strengths is a question worth investigation.

\subsection{Hamiltonian and Initial State}

We study a one-dimensional spin-$1/2$ model with local two-body interactions and onsite disorder. The Hamiltonian is given by
\be
H = H_{\text{h}}+ H_{\text{XXZ}}, 
\label{eq:H} 
\ee
where
\ba
H_{\text{h}} &=&  J\sum_{k=1}^L h_k S_k^z , \nonumber \\
H_{\text{XXZ}} &=& J\sum_{k=1}^L (S_k^x S_{k+1}^x + S_k^y S_{k+1}^y + \Delta S_k^z S_{k+1}^z).
\label{eq:Hspin}
\ea
Above, $\hbar=1$, $S_k^{x,y,z}$ are the spin operators on site $k$, $L$ is the size of the chain, which has periodic conditions, $\Delta$ is the interaction strength and $J$ sets the energy scale. We fix $\Delta = 1$, unless otherwise stated.
The Zeeman splitting on each site is $J h_k$, where $h_k$ are independent random numbers uniformly distributed in $[-h,h]$ and $h$ is the disorder strength. The total magnetization in the $z$-direction is conserved. We work in the largest subspace, which has zero total $z$-magnetization and dimension 
$D=L!/(L/2)!^2\sim \sqrt{2/\pi} (2^L/\sqrt{L})$.

The model is integrable when $h=0$. It becomes chaotic for $0<h\lesssim 1$, due to the interplay between disorder and the Ising interaction $S_k^z S_{k+1}^z$. 
It approaches a many-body localized phase as $h$ increases~\cite{SantosEscobar2004,Dukesz2009,Pal2010,Nandkishore2015,Luitz2017,Altman2018,Mace2019, Luitz2015}, which happens when the disorder is larger than a critical value, $h> h_c\sim 4$ \cite{Doggen2018, Luitz2015, Pal2010}.

We denote the eigenstates of $H_{\text{h}}$ by $|n\rangle$ and the eigenstates and eigenvalues of $H$ by  $|\alpha\rangle$ and $E_\alpha$, respectively. The initial state $\left|\Psi (0)\right>$ that we choose is an eigenstate of $H_{\text{h}}$ with energy very close to the center of the spectrum,
\be
E_0 = \langle \Psi(0)|H|\Psi(0)\rangle = \sum_\alpha\left|c_\alpha^0\right|^2 E_\alpha \sim 0,
\ee 
where $c_\alpha^0=\left<\alpha|\Psi (0)\right>$. 

In our plots, we perform averages over $0.01D$ initial states with $E_0  \sim 0$ and $10^4/(0.01D)$ disorder realizations, so that the total amount of data is $10^4$.

\subsection{$\ell$-bits Model}

Dephasing and dissipation are nonexistent in the localized phase of non-interacting systems, but in interacting systems, 
dephasing is present and responsible for the logarithmic growth of the entanglement entropy~\cite{Bardarson2012, Serbyn_2013_enta}.  Just as in the non-interacting case, all eigenstates are still localized and defined through a set of (almost) local integrals of motion $\{ \vec{\tau}_k \}$. The corresponding operators $\{ \vec{\tau}_k \}$ are adiabatically connected to the original spin operators $\{ \vec{\sigma}_k \}$ through a sequence of quasi-local unitary transformations~\cite{Imbrie2016, Huse2014, Serbyn2013} and are referred to as pseudospins or $\ell$-bits. The interaction couples the integrals of motion, giving rise to the dephasing  mechanism responsible for the entanglement growth and quantum information propagation~\cite{Bardarson2012, Serbyn_2013_enta, Tomasi2019}.

The Hamiltonian of the $\ell$-bits model describes the interacting system in the many-body localized phase and is given by
\be
H_{\ell-\text{bits}} = \sum_{k} {\cal J}_k^{(1)} \tau_k^z +  \sum_{k,l} {\cal J}_{k,l}^{(2)} \tau_k^z \tau_l^z + \ldots,
\label{Hlbit}
\ee
where $ {\cal J}_k^{(1)} $ is associated with the random fields and the coupling parameters ${\cal J}_{k,l}^{(n\geq2)}$ fall off exponentially with the distance between the sites. Building the integrals of motion of this Hamiltonian is not trivial. To circumvent the difficulties, an efficient method was proposed in Ref.~\cite{Tomasi2019}. 
The basic idea is to resort to the limit of weak interaction and strong disorder, where the higher order terms of $H_{\ell-\text{bits}}$ can be neglected and the integrals of motion of the non-interacting limit can be used. We summarize the main steps below, but for more details, see Ref.~\cite{Tomasi2019}.

Using the Jordan-Wigner transformation, the Hamiltonian $H$ in Eq.~(\ref{eq:H}) can be rewritten in terms of interacting spinless fermions
\be
H = \frac{J}{2}  \sum_{k=1}^L \left  ( c^\dagger_{k+1} c_{k} +  c^\dagger_{k} c_{k+1}  +  2 h_k \tilde{n}_k \right ) + J\Delta \sum_k^L \tilde{n}_k \tilde{n}_{k+1},
\ee
where $c^\dagger_k$ ($c_k$) is the creation (annihilation) fermionic operator at site $k$ and $\tilde{n}_k = c^\dagger_k c^\dagger_k -\frac{1}{2}$. In the weakly interacting limit ($\Delta/h \ll 1$), as a first approximation, the integrals of motion can be approximated by those of the non-interacting case. Recall that for $\Delta = 0$, the system is Anderson localized and its exact integrals of motion are given by $a^\dagger_k a_k$, where $ a_k^{\dagger} (a_k)$ 
is the creation (annihilation) operator for a single-particle Anderson eigenstate $\phi_k$ with eigenvalue $\epsilon_k$. Thus, we obtain the following effective $\ell$-bits Hamiltonian,
\be
H_{\text{eff}} = J \sum_{k} \epsilon_k a_k^{\dagger} a_k + J \sum_{l,k} S_{l,k} a_l^{\dagger} a_l a_k^{\dagger} a_k,
\label{eq:H_eff}
\ee
where $S_{l,k} = J\Delta \sum_x  [ |\phi_l(x)|^2 |\phi_k(x+1)|^2 -  \phi_l(x) \phi_l(x+1) \phi_k(x) \phi_k(x+1) ]$. Since the single-particle wavefunctions are localized, we have $S_{l,k} \sim e^{-d(l,k)/\xi}$, where $d(l,k)$ is the distance between the centers of localization of $\phi_l$ and $\phi_k$ and $\xi$ is the localization length. 

The strength of this approach relies on its efficiency, since the dynamics can be computed using free-fermion techniques. 
The computational resources to compute  the time evolution of local observables or correlation functions scale only polynomially with $L$~\cite{Tomasi2019, Wu_2019,DeTomasi2019}. Furthermore, in the limit of weak interactions, this method does not give just a qualitative description of the dynamics, but also a quantitative one, meaning that the relative error with respect to the exact dynamics is bounded in time~\cite{Tomasi2019, DeTomasi2019}.  

\subsection{Toy Model}

The onset of many-body localization was formally, under some mild assumptions about the energy spectral statistics, shown for the following Hamiltonian~\cite{Imbrie2016}, $H= J \sum_{k=1}^L h_k S_k^z + J \sum_{k=1}^L \xi_k S_k^x + J \sum_{k=1}^L j_k S_k^z  S_{k+1}^z $, where $h_k, \xi_k$ and $j_k$ are random variables, that is the model has random field, random transverse field, and random interactions.
This model was also employed recently in discussions about the existence of the localized phase in~\cite{AbaninARXIV}. Localization for this Hamiltonian becomes trivial when $j_k=0$, in which case a tensor product basis of eigenstates can be constructed. We use this limit with $\xi_k=1$,
\be
H_{\text{toy}} = J\sum_{k=1}^L h_k S_k^z + J\sum_{k=1}^L  S_k^x ,
\label{eq:H_toy}
\ee
as a toy model for the analysis of our results in the localized phase. This is partially similar to assuming that ${\cal J}_{k,l}^{(2)}=0$ in Eq.~(\ref{Hlbit}) and it is particularly convenient for studying the global quantities. Both $H_{\text{eff}}$ and $H_{\text{toy}}$ are employed in Sec.~\ref{Sec:lbit}, which is dedicated to the localized phase.

\subsection{Quantities}

We investigate the self-averaging behavior of two non-local quantities in real space, the survival probability and the inverse participation ratio, and two local experimental observables, the spin autocorrelation function and the connected spin-spin correlation function.

\subsubsection{Survival Probability} 

The survival probability gives the probability to find the initial state later in time,
\be
P_S(t)=\left|\left<\Psi (0)\right|e^{-iHt}\left|\Psi (0)\right>\right|^2 = \left| 
\sum_\alpha
\left|c_\alpha^0\right|^2 e^{-i E_{\alpha} t} \right|^2.
\label{eq:PS}
\ee
It is a non-local quantity in space and also in time. This autocorrelation function has been broadly studied since the 
beginnings of quantum mechanics~\cite{Khalfin1958,Fonda1978,Bhattacharyya1983,Ketzmerick1992,MugaBook,Torres2014PRA,Torres2014NJP,Torres2014PRE,Torres2014PRAb,Mazza2016,Torres2018,VolyaARXIV,Bera2018,Lerma2018,Reimann2016,Reimann2019, DeTomasi_2019_survival} and is now analyzed experimentally as well~\cite{SinghARXIV}.  It can be written in an integral form as
\be
P_S(t)= \left|  \int \rho_0(E) e^{-i E t} dt \right|^2, 
\ee
where
\be
\rho_0(E) = \sum_{\alpha} \left|c_\alpha^0\right|^2 \delta(E-E_{\alpha}) 
\ee
is the energy distribution of the initial state. The square of the width of $\rho_0(E) $,
\be
\Gamma^2 = \sum_{n \neq 0} |  \langle n |H| \Psi(0) \rangle |^2 ,
\label{Eq:Gamma}
\ee
depends on the number of states $|n\rangle $ directly coupled to the initial state, which is $\propto L$ for our spin model. In Eq.~(\ref{Eq:Gamma}), ``$n \neq 0$'' indicates that the sum is over all eigenstates of $H_{\text{h}}$, except for the initial state.

According to Eq.~(\ref{eq:PS}), at times beyond the saturation of the dynamics, that is for $t>t_{\text{R}}$,  the survival probability for each disorder realization fluctuates around its infinite-time average,
\be
\overline{P_S} = \sum_\alpha\left|c_\alpha^0\right|^4,
\label{Eq:SPsat}
\ee 
if the system does not have many degeneracies.

\subsubsection{Inverse Participation Ratio} 

The inverse participation ratio quantifies the spread of the initial state in the many-body Hilbert space defined by the states $|n\rangle$ \cite{Borgonovi2019}. It can be written as an out-of-time order correlator where the operators are projection operators~\cite{Borgonovi2019b}. It is given by
\be
\text{IPR}(t)=\sum_n\left|\left<n\right|e^{-iHt}\left|\Psi (0)\right>\right|^4 .
\label{eq:IPR}
\ee
At short times, the evolved state is still very close to $| \Psi(t) \rangle $ and the behavior of $\text{IPR}(t)$ is very similar to the square of the survival probability. This changes as $| \Psi(t) \rangle $ spreads over many states $|n\rangle$, not only those directly coupled with $| \Psi(0) \rangle $.

\subsubsection{Spin Autocorrelation Function} 

The spin autocorrelation function measures how close the spin configuration in the $z$-direction at a time $t$ is to the initial spin configuration,
\be
I(t)=\frac{4}{L}\sum_{k=1}^L \left<\Psi (0) \right|S^z_k e^{iHt} S^z_k e^{-iHt}\left|\Psi (0) \right>.
\label{Eq:I}
\ee
This quantity is similar to the density imbalance between even and odd sites measured in experiments with cold atoms~\cite{Schreiber2015}.

Using Eq.~(\ref{Eq:I}), we can show that at times beyond the saturation of the dynamics, $t>t_{\text{R}}$,  the  spin autocorrelation function fluctuates around the value,
\ba
\label{Eq:Isat}
\overline{I} &=& \overline{P_S} \\
&+& \frac{4}{L} \sum_{k=1}^L \langle \Psi (0)|S_k^z|  \Psi (0) \rangle 
\sum_\alpha \left|c_\alpha^0\right|^2 \sum _{n \neq 0} 
\left|c_\alpha^n\right|^2 \langle n|S_k^z|  n \rangle .
\nonumber
\ea

\subsubsection{Connected Spin-Spin Correlation Function} 

The connected spin-spin correlation function is given by
\ba
C(t)&=&\frac{4}{L}\sum_k \left[ \left<\Psi(t)\right|S_k^zS_{k+1}^z\left|\Psi(t)\right>\right.\\
&-&\left.\left<\Psi(t)\right|S_k^z\left|\Psi(t)\right>\left<\Psi(t)\right|S_{k+1}^z\left|\Psi(t)\right> \right] \nonumber
\ea
and is measured in experiments with ion traps~\cite{Richerme2014}. The initial states considered here are non-correlated product states in the $z$-direction, so $C(0)=0$. As the system evolves, $C(t)$ quantifies the average growth of correlations between neighboring sites.

\section{Survival Probability}
\label{Sec:SP}
In the chaotic regime, the survival probability is not self-averaging at any time scale~\cite{Schiulaz2020}. This was shown analytically by evolving $P_S(t)$ with full random matrices. Based on numerical results for all times and  analytical results for short and long times, we verified that the same is true also for the disordered chaotic spin model~\cite{Schiulaz2020}. As we now show, the survival probability remains non-self-averaging at all times as the disorder strength increases and the system approaches localization, but differences exist. One worth pointing out is that after saturation, while the relative variance of $P_S(t)$ is constant in the chaotic regime, specifically ${\cal R}_{P_S}  (t>t_{\text{R}}) \sim 1$, it grows with $L$ in the localized phase. 

\subsection{Short Times: $\mathbf{t< \Gamma^{-1}}$}
\begin{figure}[h!]
\includegraphics*[width=0.47\textwidth]{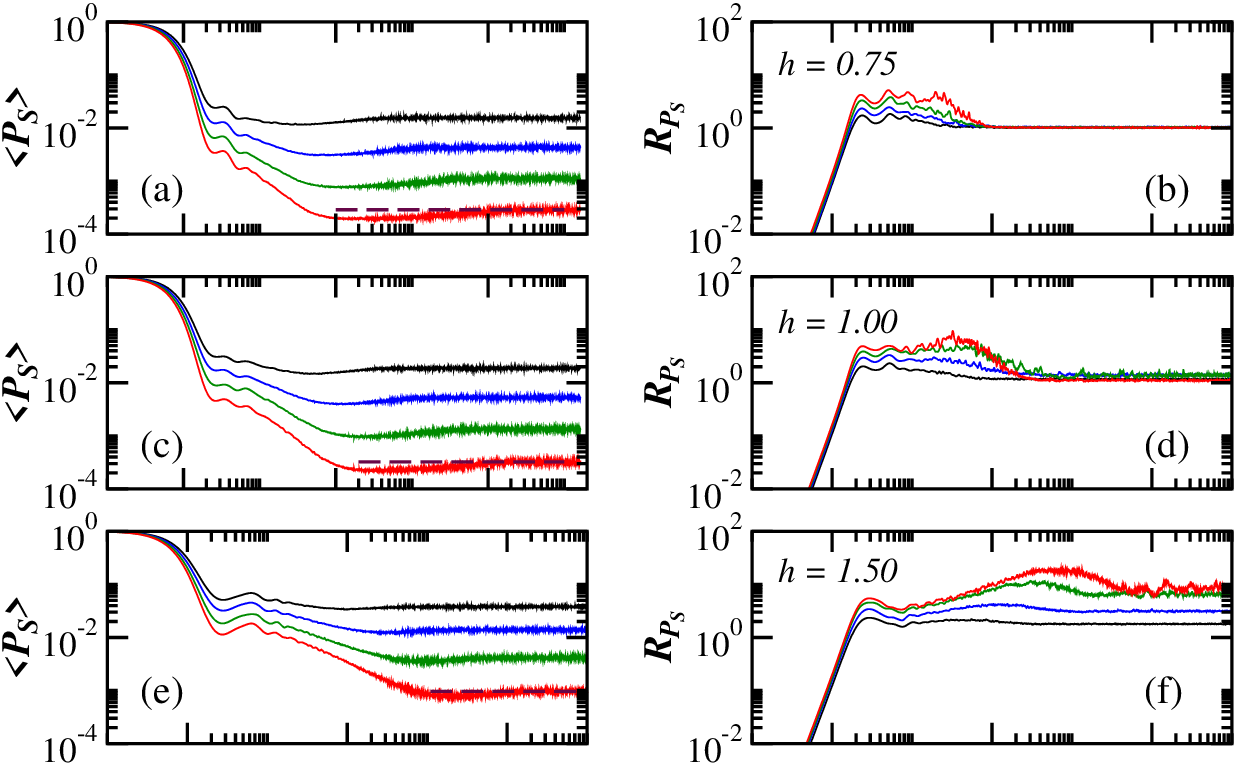}\\
\includegraphics*[width=0.47\textwidth]{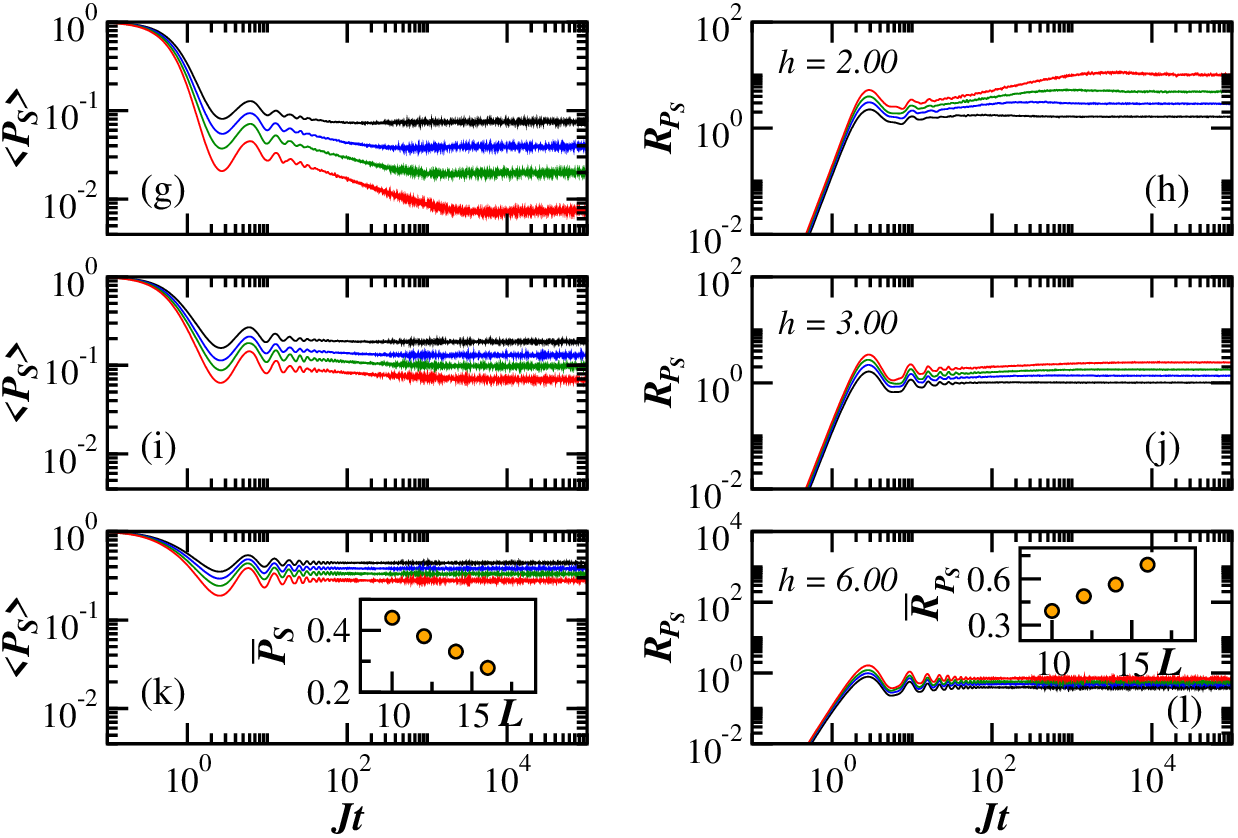}
\caption{Left panels: Mean value of the survival probability. Right panels: Relative variance of the survival probability. The values of the disorder strength are indicated on the panels, they increase from the top to the bottom panel. The curves correspond to system sizes $L=10$ (black), 12 (blue), 14 (green), and 16 (red).  The horizontal dashed line in Fig.~\ref{fig:SP}~(a), (c), and (e) indicates the saturation point, $\sum_\alpha\left|c_\alpha^0\right|^4$ [Eq.~(\ref{Eq:SPsat})],  for $L=16$.  
}
\label{fig:SP}
\end{figure}

The expansion for short times gives ${\cal R}_{P_S}(t< \Gamma^{-1})=\sigma^2_{\Gamma^2}t^4+{\cal O}(t^6)$, where $\sigma^2_{\Gamma^2}=\left<\Gamma^4\right>-\left<\Gamma^2\right>^2 $. This result is independent of the disorder strength, because according to Eq.~(\ref{Eq:Gamma}),  $\Gamma^2$ depends only on the off-diagonal elements of $H$ written in the product states, while disorder enters in the diagonal elements. This implies that the relative variance of $P_S$ increases linearly with system size for any (reasonable value of the) disorder strength,
\be
{\cal R}_{P_S} (t< \Gamma^{-1}) \propto J^4 t^4 L\,.
\label{SP_short}
\ee
This is indeed what we see in Fig.~\ref{fig:SP}, where we show the mean of the survival probability on the left panels and its relative variance on the right panels for six values of the disorder strength, from the top to the bottom panel: $h=0.75, 1, 1.5, 2, 3$ and $6$.  The value $h=0.75$ represents the chaotic region and it was studied in~\cite{Schiulaz2020}. For $h=6$, the system is already in the localized phase. There is no difference in the behavior of ${\cal R}_{P_S}(t)$ at short times for the different disorder strengths.

\subsection{Long Times: $\mathbf{t> t_{\text{R}}}$}
\label{SecSpLongT}

To compute the relative variance after saturation, ${\cal R}_{P_S}  (t>t_{\text{R}})$, we need $\left< P_S (t>t_{\text{R}})\right>$ and $\left< P_S^2(t>t_{\text{R}}) \right>$. At long times, the first term on the right hand side of the equation
\[
\left< P_S (t>t_{\text{R}})\right>=
\left< \sum_{\alpha \neq \beta}
|c_\alpha^0|^2 |c_\beta^0|^2 e^{-i (E_{\alpha} - E_{\beta}) t} \right>+  \left< \sum_\alpha\left|c_\alpha^0\right|^4 \right>
\]
cancels out in the absence of many degeneracies, so 
\be
\left< P_S (t>t_{\text{R}})\right>= \left< \sum_\alpha\left|c_\alpha^0\right|^4 \right>.
\ee

In the equation for
\ba
\left< P_S^2(t>t_{\text{R}}) \right> &=&  \nonumber \\
&& \hspace{-1.5 cm} \left< 
\sum_{\alpha, \beta, \gamma,\delta} \!\!
|c_\alpha^0|^2 \! |c_\beta^0|^2 \! |c_\gamma^0|^2  \! |c_\delta^0|^2
\! e^{-i (E_{\alpha} - E_{\beta}  + E_{\gamma} - E_{\delta}) t}  \right> \nonumber,
\ea
the terms that do not average out are $\alpha=\beta, \gamma=\delta, \alpha \neq \delta$, also $\alpha=\delta, \beta=\gamma, \alpha \neq \beta$, and $\alpha=\beta=\gamma=\delta$. Therefore,
\ba
\left< P_S^2(t>t_{\text{R}}) \right> &=& 
2 \left<  \sum_{\alpha \neq \beta} \left|c_\alpha^0\right|^4 \left|c_\beta^0\right|^4 \right> + \left< \sum_{\alpha} \left|c_\alpha^0\right|^8  \right>
\nonumber 
\ea
and
\ba
\label{Eq:R_SP}
&&{\cal R}_{P_S}  (t>t_{\text{R}})  =  \\
&&\frac{
2 \left< \left( \sum_{\alpha} \left|c_\alpha^0\right|^4  \right)^2 \right>
-  \left< \sum_{\alpha} \left|c_\alpha^0\right|^4  \right> ^2
 - \left< \sum_{\alpha} \left|c_\alpha^0\right|^8 \right>
}{\left< \sum_{\alpha} \left|c_\alpha^0\right|^4  \right> ^2}  .
\nonumber
\ea

In the chaotic regime, the eigenstates away from the edges of the spectrum, and thus also our initial states, are similar to the eigenstates from full random matrices, that is, they are approximately normalized random vectors. This means that the coefficients $c_\alpha^0 $ are nearly random numbers from a Gaussian distribution with the constraint $ \sum_{\alpha} \left|c_\alpha^0\right|^2 =1$, in other words, $\left|c_\alpha^0\right|^2 \sim 1/D$. This implies that $\left< P_S (t>t_{\text{R}})\right> \propto 1/D$, as seen in Fig.~\ref{fig:SP}~(a). Due to the uniformization of the components of the initial state,
\be
{\cal R}_{P_S}  (t>t_{\text{R}}) \simeq 1,
\ee
as seen in Fig.~\ref{fig:SP}~(b). This implies that the long-time relative variance of  the survival probability is independent of the system size and this quantity is thus non-self-averaging. 

As the disorder strength increases above 1, $\left< P_S (t>t_{\text{R}})\right> $ grows and ${\cal R}_{P_S}  (t>t_{\text{R}})$  becomes dependent on the system size, reaching values even larger than 1, as e.g. for $h=2$ in Fig.~\ref{fig:SP}~(h). 
This is expected, since by increasing $h$ above 1, the eigenstates distance themselves from those of full random matrices, correlations build up between their components, and thus the fluctuations of $P_S (t)$ at long times should increase. At first sight, these results suggest the onset of multifractal eigenstates~\cite{Luca2014,Torres2015,Torres2017,Pino2017}, meaning that they do not span homogeneously the entire Hilbert space, but only a vanishing portion of the the full Hilbert space, so  $\left< P_S (t>t_{\text{R}})\right> \propto D^{-\gamma}$ with $0< \gamma<1$.  However, multifractality at intermediate disorder strengths, $1<h<h_c$, has been challenged in Ref.~\cite{Mace2019}, where system sizes up to $L=24$ were considered. 

In the many-body localized phase, on the other hand, it has been argued that the eigenstates are indeed multifractal~\cite{Mace2019, Luitz2015}. This should imply $\left< P_S (t>t_{\text{R}})\right> \propto D^{-\gamma}$ and the exponential growth of ${\cal R}_{P_S}  (t>t_{\text{R}})$ with $L$. With the few numerical points in the insets of Fig.~\ref{fig:SP}~(k) and Fig.~\ref{fig:SP}~(l), it is not possible to confirm that. In fact, one cannot even exclude linear scalings with $L$,  that is $\left< P_S (t>t_{\text{R}})\right> \propto 1/L$ and ${\cal R}_{P_S}  (t>t_{\text{R}}) \propto L$. 
Results better aligned with the expectation of multifractality are obtained with the toy model (\ref{eq:H_toy}), as discussed in Sec.~\ref{Sec:lbit}.

Independently of the region where multifractality emerges and on the proper scaling of $\left< P_S (t>t_{\text{R}})\right>$ and ${\cal R}_{P_S}  (t>t_{\text{R}})$, it is unquestionable that in the localized phase, the survival probability after saturation remains non-self-averaging, but now at an even stronger sense that in the chaotic regime.

\subsection{Intermediate Times: $ \mathbf{\Gamma^{-1}<t<t_{\text{R}}}$}

At intermediate times, $\Gamma^{-1}<t<t_{\text{R}}$, one sees that the oscillations observed in the evolution of $\left< P_S (t) \right> $ get reflected in oscillations for ${\cal R}_{P_S}(t) $ as well. The envelope of the oscillations of $\left< P_S (t) \right> $ follow a power-law decay~\cite{Tavora2016,Tavora2017}. In the chaotic region, this power-law behavior is in part caused by the presence of the edges of the spectrum~\cite{Tavora2016,Tavora2017, DeTomasi_2019_survival}, where the eigenstates are not chaotic. Beyond chaos, the power-law decay is caused by correlations between the components of the eigenstates~ \cite{Torres2015,Torres2017}. The absence of chaotic states in both scenarios is a possible justification for the values of ${\cal R}_{P_S}  (t)$ above 1 seen for times $t\sim10 J^{-1}$.

Another interesting feature appears after the power-law decay of  $\left< P_S (t) \right> $ and before saturation. When the eigenvalues have some degree of correlation, be it in the chaotic regime or in the intermediate region between chaos and localization, $\left< P_S (t) \right> $ shows a dip below the saturation point $\left< \overline{P_S} \right>$, which is known as correlation hole~\cite{Leviandier1986}. In Figs.~\ref{fig:SP}~(a), (c), and (e), the dip is clearly seen below the horizontal dashed line that marks $\left< \overline{P_S} \right>$. For reasons explained in Ref.~\cite{Schiulaz2019}, we call ``Thouless time'' \cite{footSierant} the time to reach the minimum of the correlation hole and denote it by $t_{\text{Th}}$. This is the point where the dynamics resolve the discreteness of the spectrum and detect spectral correlations.

The correlation hole becomes less deep~\cite{Torres2017Philo} and $t_{\text{Th}}$ is postponed to longer times~\cite{Schiulaz2019} as $h$ increases and the correlations between the eigenvalues die off [cf. Figs.~\ref{fig:SP}~(a), (c), and (e)]. In the chaotic regime, there is no difference in the behavior of ${\cal R}_{P_S}(t) $ during or after the correlation hole [cf. Fig.~\ref{fig:SP}~(a) and Fig.~\ref{fig:SP}~(b)]. During this entire interval, $t_{\text{Th}}<t<t_{\text{R}}$, the dynamics depend only on the eigenvalues and no longer on the components of the initial state. However, in the intermediate regime, such as for $h=1.5$, the behavior of ${\cal R}_{P_S}(t) $ for $t\sim t_{\text{Th}}$ is different from that for $t>t_{\text{R}}$ [cf. Figs.~\ref{fig:SP}~(e) and Figs.~\ref{fig:SP}~(f)]. In the region of the hole, ${\cal R}_{P_S}(t)$ is pushed to larges values, while  for $t>t_{\text{R}}$, ${\cal R}_{P_S}(t)$ saturates at a lower point. It is likely that for $h\sim 1.5$, the dependence of the dynamics on the components $|c_{\alpha}^0|^2$ of the initial state persists to later times, including the interval of the correlation hole, fading away only when saturation is approached.

\section{Inverse Participation Ratio}
\label{Sec:IPR}

When the system is chaotic, the inverse participation ratio exhibits two different behaviors. It is non-self-averaging at short times, but becomes self-averaging at long times~\cite{Schiulaz2020}, once the initial state has had time to spread in the many-body Hilbert space and to visit the exponentially large number of many-body states accessible to its energy. In contrast, as shown below, as we approach the localized phase, this global quantity becomes non-self-averaging at all times. 

The dependence on the system size  of ${\cal R}_{\text{IPR}}(t)$ at long times clearly distinguishes chaos from localization. In the ergodic phase, where the eigenstates are close to random vectors, ${\cal R}_{\text{IPR}}(t>t_{\text{R}}) \propto 1/D$.  Contrary to that, in the localized phase, ${\cal R}_{\text{IPR}}(t>t_{\text{R}})$ increases as $L$ grows.

\subsection{Short Times: $\mathbf{t< \Gamma^{-1}}$}

At short times, the evolved state $|\Psi(t)\rangle$ is not yet very far from $|\Psi(0)\rangle$, so the inverse participation ratio behaves similarly to the square of the survival probability and Eq.~(\ref{SP_short}) applies also for ${\cal R}_{\text{IPR}}(t<\Gamma^{-1})$. Accordingly, as shown on the right panels of Fig.~\ref{fig:IPR}, ${\cal R}_{\text{IPR}}(t)$ increases linearly with $L$ independently of the disorder strengths considered.

\subsection{Long Times: $\mathbf{t> t_{\text{R}}}$}

\begin{figure}[htb]
\includegraphics*[width=0.47\textwidth]{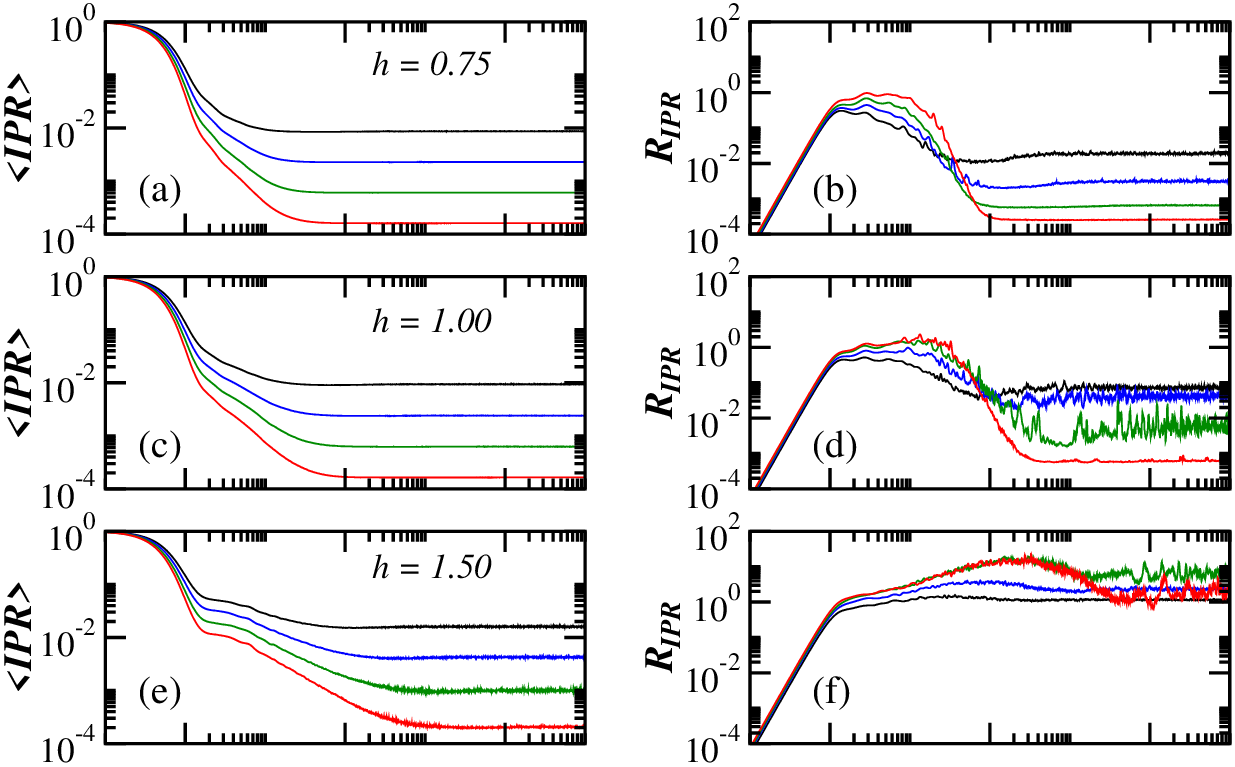}\\
\includegraphics*[width=0.47\textwidth]{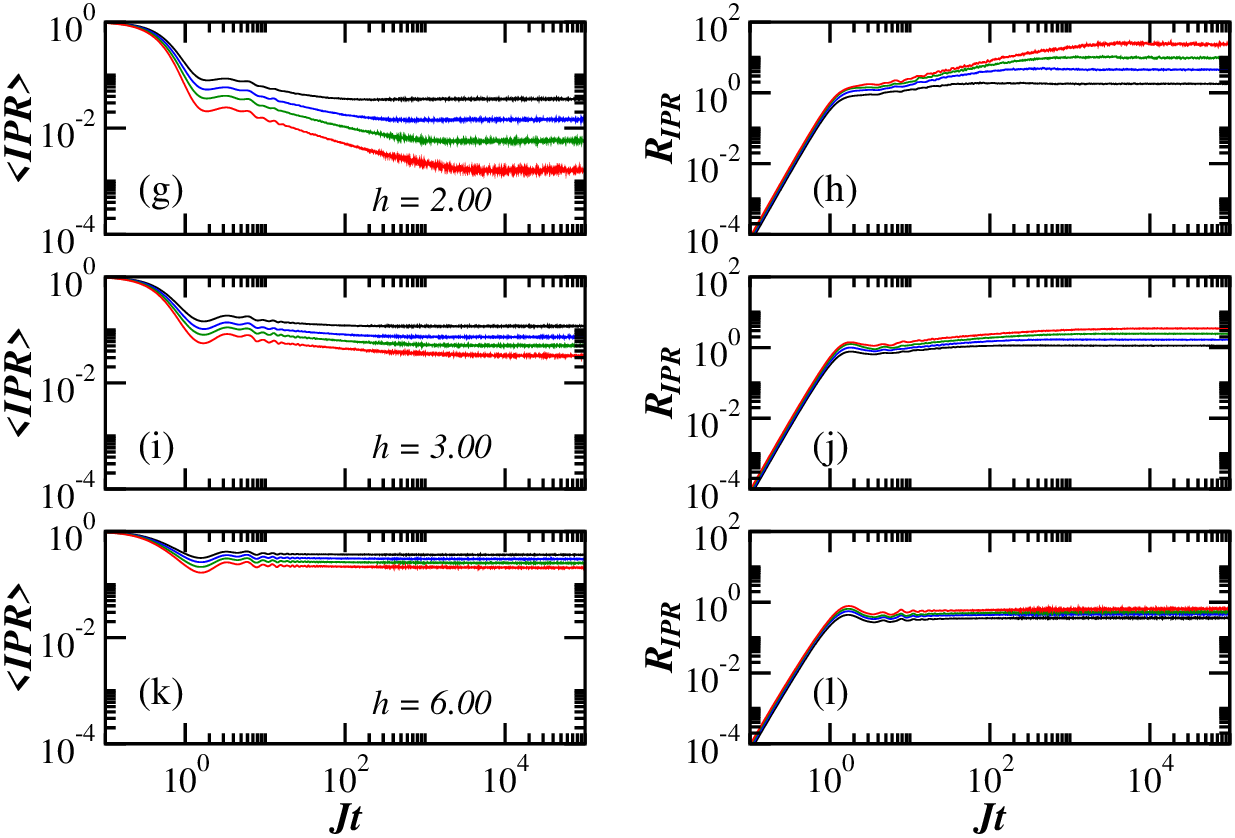}
\caption{Left panels: Mean value of the inverse participation ratio. Right panels: Relative variance of the inverse participation ratio. The values of the disorder strength are indicated on the left panels, they increase from the top to the bottom panel. The curves correspond to system sizes $L=10$ (black), 12 (blue), 14 (green), and 16 (red). }
\label{fig:IPR}
\end{figure}

To study ${\cal R}_{\text{IPR}}(t)$ at times $t>t_{\text{R}}$, we write the inverse participation ratio  in terms of the energy eigenstates,
\ba
\text{IPR}(t)&=&\sum_n\sum_{\alpha,\beta,\gamma,\delta}e^{-i(E_\alpha-E_\beta+E_\gamma-E_\delta)t}\nonumber\\
&\times& c_\alpha^0 c_\alpha^{n*} c_\beta^n c_\beta^{0*} c_\gamma^0 c_\gamma^{n*} c_\delta^n c_\delta^{0*} ,
\label{eq:IPRfull}
\ea
and use the same reasoning employed in the analysis of ${\cal R}_{P_S}(t>t_{\text{R}})$. The difference now is that one has an additional sum over all unperturbed many-body states $|n\rangle$. In the chaotic regime, this significantly reduces the fluctuations and leads to ``super'' strong self-averaging, that is, the relative variance of $\text{IPR}(t)$ decreases exponentially with $L$. As seen in Fig.~\ref{fig:IPR}~(b), ${\cal R}_{\text{IPR}}(t>t_{\text{R}}) \propto 1/D$, which can be explained analytically using the fact that the eigenstates are nearly random vectors~\cite{TorresARXIV}.

The inverse participation ratio does not develop a visible correlation hole [cf. Fig.~\ref{fig:SP}~(a) and Fig.~\ref{fig:IPR}~(a)]. The hole exists, with its minimum at the same time $t_{\text{Th}}$ \cite{Schiulaz2020}, but it is minor and the ratio between the minimum value of $\text{IPR}(t)$ and it saturation value goes to 1 exponentially fast as $L$ increases. In spite of that, the behavior of ${\cal R}_{\text{IPR}}(t) $ before and after the hole is clearly different in the chaotic region.

The picture above changes as $h$ increases above 0.75  and the minimum of the correlation hole moves to longer times~\cite{Torres2017Philo,Schiulaz2019}. The crossings between the curves for ${\cal R}_{\text{IPR}}(t)$ happen now at ever longer times [cf. Fig.~\ref{fig:IPR}~(b), (d), and (f)]. The instant where the spectral correlations get dynamically detected marks the time beyond which $\text{IPR}(t)$ becomes self-averaging. This point disappears once the correlations are destroyed. For disorder strengths $h>1.5$, the curves for the system sizes considered here no longer cross and the inverse participation ratio becomes non-self-averaging at all times. 

Using the same arguments of multifractality in the many-body localized phase discussed in Sec.~\ref{SecSpLongT}, we find that $\text{IPR}(t>t_{\text{R}}) \propto e^{-\gamma L}$ and ${\cal R}_{\text{IPR}}(t>t_{\text{R}}) \propto e^{\gamma L}$, as explained in Sec.~\ref{Sec:lbit}. In fact, according to the toy model (\ref{eq:H_toy}), both ${\cal R}_{P_S}(t)$ and ${\cal R}_{\text{IPR}}(t)$ grow exponentially with $L$ at all times. 

In the next two sections, we contrast the self-averaging properties of the non-local quantities described in Sec.~\ref{Sec:SP} and Sec.~\ref{Sec:IPR} with those for the local quantities.

\section{Spin Autocorrelation Function}

The self-averaging behavior of the spin autocorrelation function with respect to time is just the opposite from the inverse participation ratio. $I(t)$ is  self-averaging at short times for any value of the disorder strength, while at long times, whether self-averaging holds or not depends on the disorder strength. In the chaotic regime, just as the survival probability and contrary to the inverse participation ratio, the spin autocorrelation function is non-self-averaging for $t>t_{\text{Th}}$. As the disorder strength increases, the non-self-averaging region is pushed to ever longer times, until the system reaches the localized phase, where $I(t)$ becomes self-averaging at all times.

\subsection{Short Times: $\mathbf{t< \Gamma^{-1}}$}

\begin{figure}[htb]
\includegraphics*[width=0.47\textwidth]{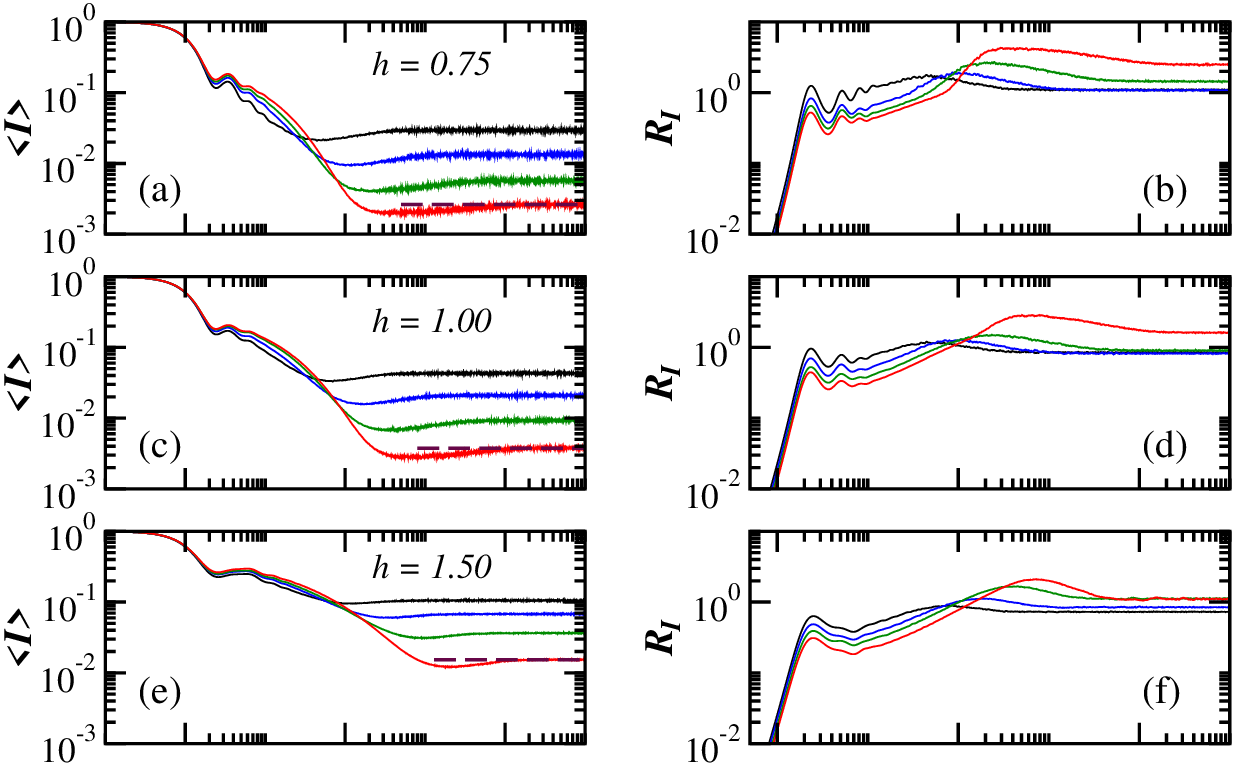}\\
\includegraphics*[width=0.47\textwidth]{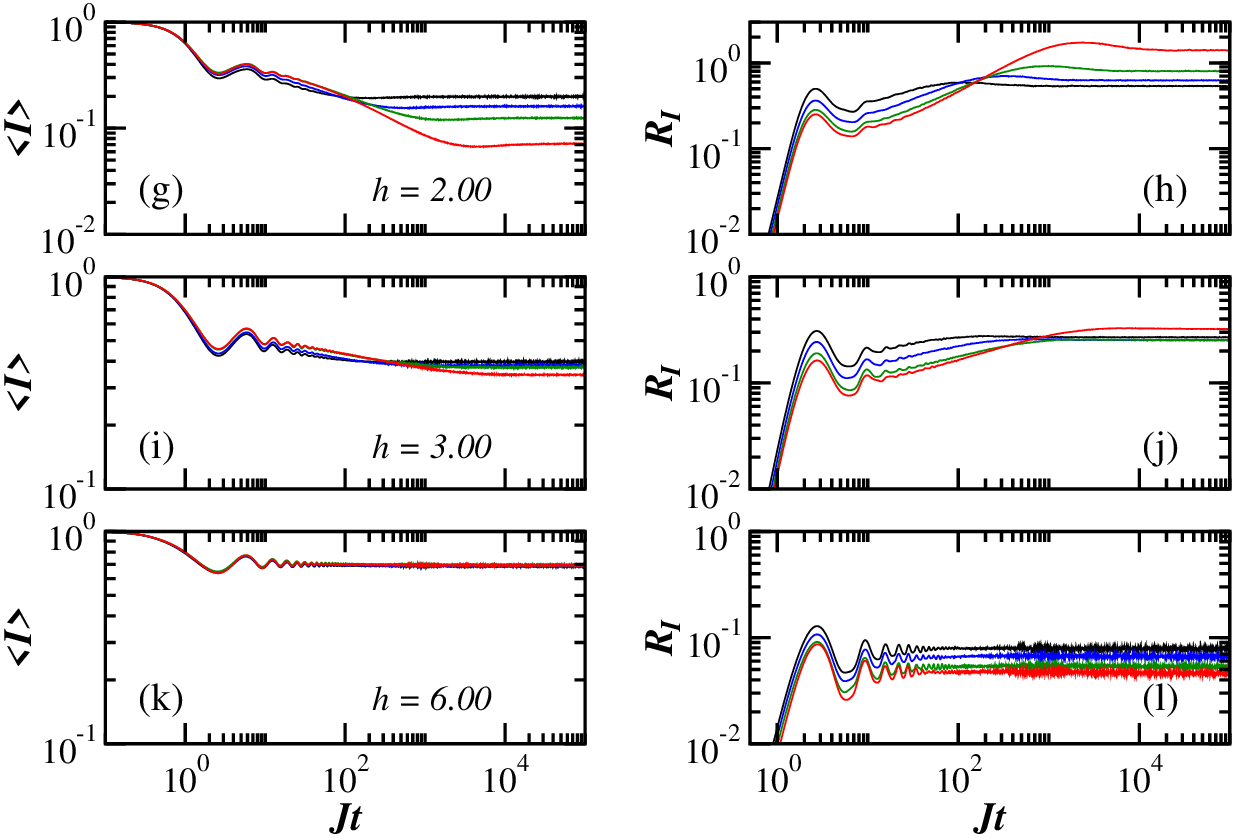}
\caption{Left panels: Mean value of the spin autocorrelation function. Right panels: Relative variance of the spin autocorrelation function. The values of the disorder are indicated on the left panels, they increase from the top to the bottom panel. The curves correspond to system sizes $L=10$ (black), 12 (blue), 14 (green), and 16 (red). The horizontal dashed line in Fig.~\ref{fig:I}~(a), (c), and (e) marks the saturation value of the spin autocorrelation function [Eq.~(\ref{Eq:Isat})] for $L=16$. 
}
\label{fig:I}
\end{figure}

Quantities that are local in space and involve spatial averages, such as the spin autocorrelation function, are self-averaging at short times for any value of the disorder strength, which can be understood as follows. For $t<\Gamma^{-1}$, the excitations only have time to hop to few neighboring sites, even if the system is deep in the chaotic phase. As a result, due to the spatial average, which corresponds to the sum over $k$ in Eq.~(\ref{Eq:I}), the relative variance decreases with system size. This can be seen by expanding the relative variance of the spin autocorrelation function for short times~\cite{Schiulaz2020}, which gives
\be
{\cal R}_I(t) = \frac{16 \sigma_{\Gamma^2}^2 t^4}{L^2}+{\cal O}(t^6) \propto \frac{J^4t^4}{L},
\ee
where $1/L$ appears explicitly. Since $\sigma_{\Gamma^2}$ does not depend on $h$, the above result is independent of the disorder strength. This statement is confirmed by the right column of Fig.~\ref{fig:I}, where the short-time behavior of ${\cal R}_I(t<\Gamma^{-1})$ is very similar from Fig.~\ref{fig:I}~(b) to Fig.~\ref{fig:I}~(l).

\subsection{Long Times: $\mathbf{t> t_{\text{R}}}$}

As seen in Fig.~\ref{fig:I}~(b), the curves for ${\cal R}_I(t)$ in the chaotic regime cross in the region of the correlation hole. Similarly to the survival probability, the spin autocorrelation function is not self-averaging at long times. $P_S(t)$ and $I(t)$ are both autocorrelation functions, which may explain some of their common features.  The two quantities develop a visible correlation hole, as seen for $I(t)$ in Figs.~\ref{fig:I}~(a), (c), and (e). We note, however, that contrary to $P_S(t)$, our numerical studies (not shown) indicate that the correlation hole for $I(t)$ shrinks for $L>16$. It is an open question whether the onset of a visible correlation hole has any direct connection with the lack of self-averaging at long times in the chaotic regime.

As $h$ increases above 0.75 and the correlation hole gets postponed, one should expect the crossings between the curves of ${\cal R}_I(t)$ to happen later in time, analogously to what one sees for the inverse participation ratio. This is indeed the case, but  for the spin autocorrelation function it only becomes evident for $h>1.5$. Despite the shift of the correlation hole to longer times in Figs.~\ref{fig:I}~(a) and (c), the crossing of the curves of ${\cal R}_I(t)$ in Figs.~\ref{fig:I}~(b) and (d) happens at similar times. It is only for $h>1.5$ that we finally see a clear shift in the crossing points of ${\cal R}_I(t)$ to later times. It may be that the spin autocorrelation function is more sensitive to finite size effects. 

In the many-body localized phase, $h=6$, the spin autocorrelation function becomes self-averaging also at long times, as seen in Fig.~\ref{fig:I}~(l). This reflects the locality of the observable. For $h>h_c$,  the initial spin configuration cannot change much in time, as clearly seen in Fig.~\ref{fig:I}~(k). In contrast to the cases with $h<h_c$, $\langle I(t>t_{\text{R}}) \rangle$ in Fig.~\ref{fig:I}~(k) no longer depends on the system size and saturates at a finite value that does not decrease with $L$. This behavior contrasts also with that for the non-local quantities, where even in the localized phase, one sees that $ \langle P_S(t>t_{\text{R}}) \rangle$ and $\langle \text{IPR}(t>t_{\text{R}}) \rangle$ depend on $L$ [Fig.~\ref{fig:SP}~(k) and Fig.~\ref{fig:IPR}~(k)]. The fact that the values of $I(t>t_{\text{R}})$ do not depend on $L$ explain why, as the system size increases, the variance $\sigma_I^2(t)$ can still decrease with $L$, resulting in the self-averaging behavior of $I(t)$. 

A more quantitative analysis of the self-averaging behavior of the spin autocorrelation function at all times in the many-body localized phase is provided in Sec.~\ref{Sec:lbit}.

\section{Connected Spin-Spin Correlation Function}

The connected spin-spin correlation function combines all the good properties for self-averaging found in the previous quantities. It is local in space, as the spin autocorrelation function, so it is self-averaging at short times for any disorder strength. It is not an autocorrelation function and does not develop a correlation hole, which may explain why it is self-averaging also at long times for any value of $h$. This quantity is thus self-averaging at any time scale and for any disorder strength, which is the perfect picture for an experimental quantity. 

\begin{figure}[ht]
\includegraphics*[width=0.47\textwidth]{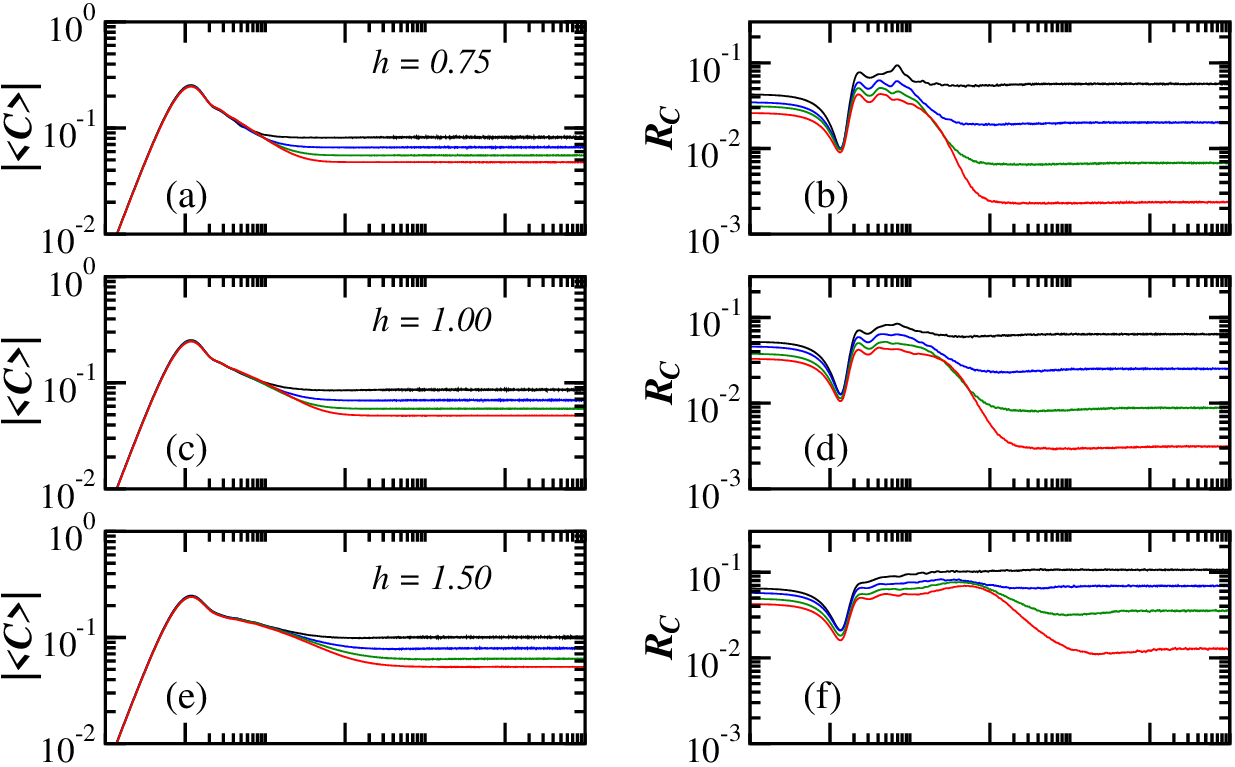}\\
\includegraphics*[width=0.47\textwidth]{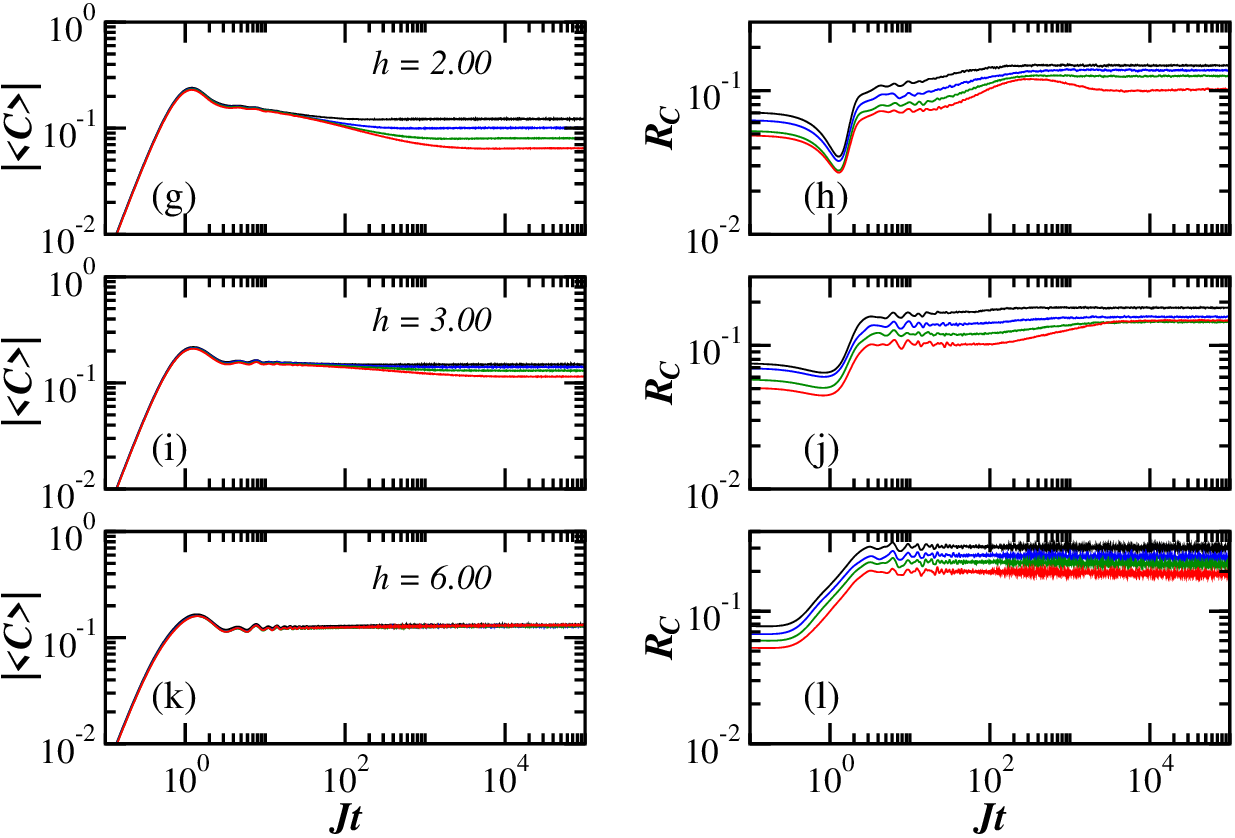}
\caption{Left panels: Mean value of the connected spin-spin correlation function. Right panels: Relative variance of the connected spin-spin correlation function. The values of the disorder are indicated on the left panels, they increase from the top to the bottom panel. The curves correspond to system sizes $L=10$ (black), 12 (blue), 14 (green), and 16 (red).
}
\label{fig:C}
\end{figure} 

In Fig.~\ref{fig:C}, we show the absolute value of the mean of $C(t)$ on the left columns and the relative variance of $C(t)$ on the right columns, confirming its self-averaging behavior for all $h$'s and all times. But as seen in the panels, the values of ${\cal R}_C(t)$ depend on the time scale and on the disorder strength.

The relative variance of $C(t)$ has a non-monotonic behavior in time in the chaotic and intermediate regimes, showing a dip at $t\sim 1$ and a bump at $t\sim t_{\text{Th}}$. This reflects the behavior of the mean of the spin-spin correlation function, which is very fast for $t<\Gamma^{-1}$, but then slows down up to $t_{\text{Th}}$. 
The slow dynamics in the interval $[\Gamma^{-1},t_{\text{Th}}]$ is observed for all four quantities. This is the time of the power-law decay of the survival probability [see Fig.~\ref{fig:C} (a), (c), (e)]. The interval gets elongated as the disorder strength increases and ${\cal R}_C(t)$ thus takes longer to saturate [cf. Figs.~\ref{fig:C}~(b), (d), (f), (h), (j)]. 

The non-monotonic behavior of ${\cal R}_C(t)$ disappears in the localized phase [Fig.~\ref{fig:C}~(l)], where the initial fast evolution of $|\langle C(t) \rangle |$ is simply followed by the saturation of the dynamics [Fig.~\ref{fig:C}~(k)]. In this phase, similarly to what was seen for the spin autocorrelation function in Fig.~\ref{fig:I}~(k), and contrary to the behavior of the global quantities, the values  of $C(t)$ are independent on the system size.

At equilibrium, the scaling of ${\cal R}_C(t>t_{\text{R}})$ with system size makes evident the difference between the chaotic and the many-body localized phase. In the chaotic regime, the relative fluctuations of the spin-spin correlation decrease exponentially with system size, ${\cal R}_C(t>t_{\text{R}}) \propto 1/D$, while in the localized phase, it decreases linearly with $L$, as justified with the $\ell$-bits model in the next section.

\section{Localized Phase}
\label{Sec:lbit}

This section is dedicated to the limit of strong disorder, $h=6$, which is already deep in the localized phase. With the noninteracting model ($\Delta=0$) from Eq.~(\ref{eq:H}),  the $\ell$-bits Hamiltonian in Eq.~(\ref{eq:H_eff}), and the toy model in Eq.~(\ref{eq:H_toy}), we can inspect large systems size and asymptotically long times. Notice that the plots in this section are semi-log, while the plots in previous sections were log-log.

\subsection{Global Quantities}

We start by discussing the global quantities. We perform the scaling analysis of the survival probability using the noninteracting model in Eq.~(\ref{eq:H}) with $\Delta=0$ and explain the results with the toy model in Eq.~(\ref{eq:H_toy}). 

\subsubsection{Survival Probability}

\begin{figure}[h!]
\includegraphics*[width=1.\linewidth]{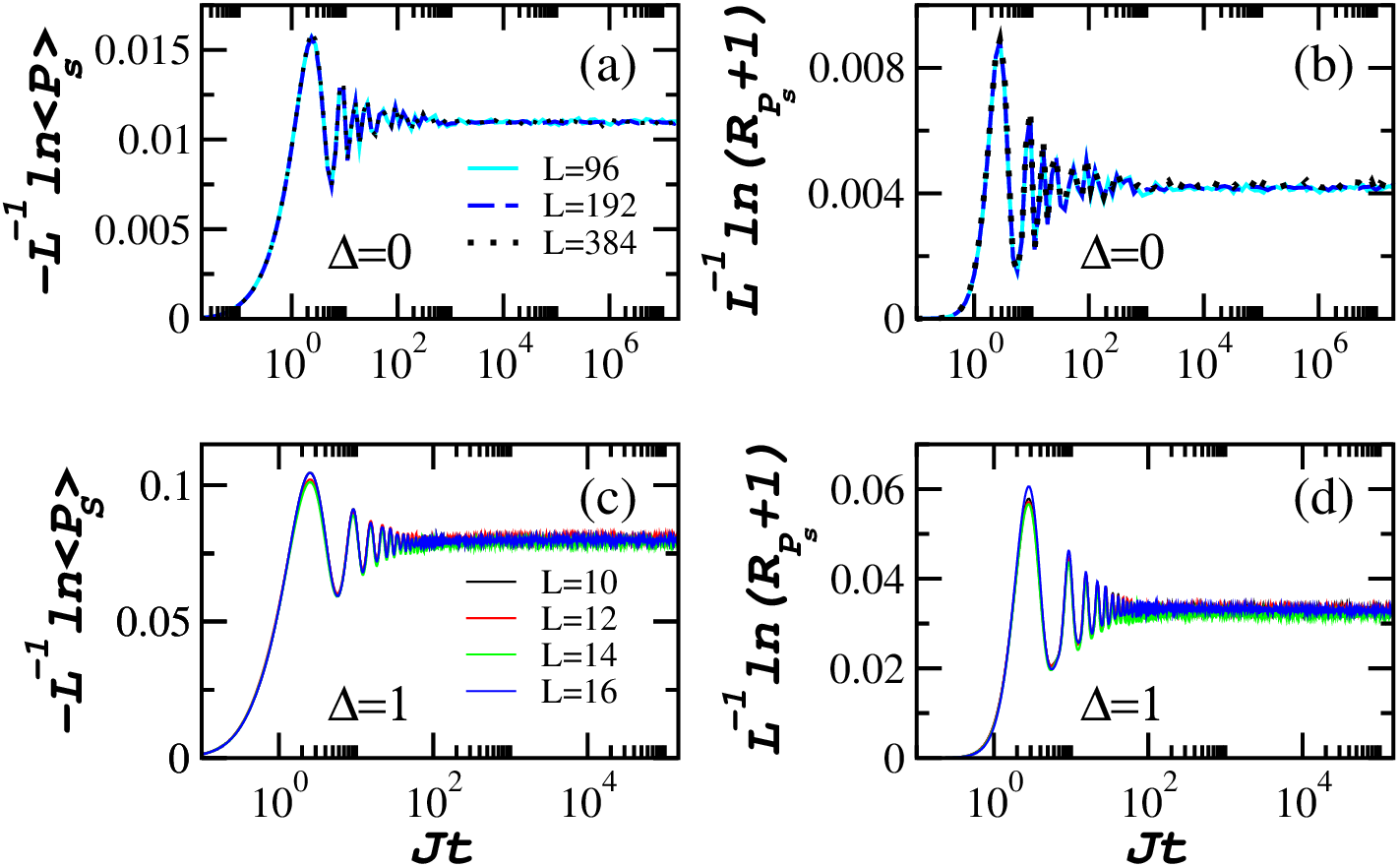}
\caption{Rescaled survival probability $-L^{-1} \ln \langle P_S\rangle $  in (a) and (c) and its rescaled relative variance $-L^{-1} \ln(\langle P_S\rangle +1)$ in (b) and (d) for the noninteracting model $(\Delta=0)$ of Eq.~(\ref{eq:H}) in (a) and (b) and for the interacting case $(\Delta=1)$ in (c) and (d). The system sizes are indicated.}
\label{fig:Return_non}
\end{figure}

To try to understand the dependence of $\langle P_S(t) \rangle$ and $\mathcal{R}_{P_S}(t)$ on system size $L$, we consider the noninteracting limit, $\Delta = 0$, of the Hamiltonian in Eq.~(\ref{eq:H}). Figures~\ref{fig:Return_non}~(a) and (b) display $- L^{-1} \ln \langle P_S(t)\rangle $ and $L^{-1} \ln( \mathcal{R}_{P_S}(t)+1)$ for $h=6$, $\Delta=0$, and large system sizes. The plots show that the survival probability is exponentially suppressed in system size, $P_S(t) \propto e^{-\gamma L}$, while $\mathcal{R}_{P_S}(t)$  increases exponentially fast with system size,  $\mathcal{R}_{P_S}(t) \propto e^{\gamma L}$. The exponential decay of the survival probability with system size after equilibration implies that $\sum|c_{\alpha}^0|^4 \propto {\cal D}^{-\gamma}$ with $0<\gamma<1$, which agrees with the picture of fractal eigenstates  in the many-body localized phase~\cite{Mace2019}.

The equivalent plots in Figs.~\ref{fig:Return_non}~(c) and (d) are obtained with the numerical data from Fig.~\ref{fig:SP}~(k)-(l) for the interacting Hamiltonian of Eq.~(\ref{eq:H}). The top and bottom panels of Fig.~\ref{fig:Return_non} are very similar indicating that the limit $\Delta=0$ and the scaling analysis obtained with it describe well the behavior of the survival probability and its fluctuations in the many-body localized phase. 

We verified that if we consider $ -\langle \ln{P_S}(t) \rangle $ instead of $-\ln \langle P_S(t) \rangle$, we recover self-averaging, that is $\mathcal{R}_{-\ln{P_S}}(t) \propto L^{-1}$ (not shown). This comes from the fact that logarithms cut the tails of distributions, thus favoring self-averaging. A somewhat similar discussion appears in Ref.~\cite{Schiulaz2020} when comparing the self-averaging behavior of $\text{IPR}(t)$ and of the second-order R\'enyi entropy $-\ln[\text{IPR}(t)]$.

To provide an explanation of the behavior of the survival probability and its fluctuations we employ the toy model in Eq.~(\ref{eq:H_toy}). In this case, it is straightforward to compute the survival probability
\be
P_S(t) = \prod_k^L f_k(t),
\label{eq:prod}
\ee
with
\begin{multline}
f_k(t)=  \cos^4 ({\phi_k/2}) + \sin^4 ({\phi_k/2}) +\\
2\cos^2 ({\phi_k/2}) \sin^2 ({\phi_k/2})  \cos{[(\epsilon_+^{(k)}-\epsilon_-^{(k)})t]},
\label{eq:Sur_non_interacting}
\end{multline}
where $\sin^2 \phi_k = 1/\sqrt{1+h_k^2}$ and $\epsilon_{\pm}^{(k)} = \pm \sqrt{1+h_k^2}$. 
For a fixed time $t>0$, the functions $\{f_k(t)\}$ are positive independently and identically distributed random variables with mean value $\langle f_k(t) \rangle \sim e^{-\gamma(t)}$ and second moment $\langle f_k^2(t) \rangle \sim e^{-\gamma_1(t)}$, where $\gamma(t), \gamma_1(t)>0$.
Thus, $\langle P_S(t) \rangle = \prod^L \langle f_k(t) \rangle = e^{-\gamma(t)L}$ is exponentially suppressed in system size. 
Instead, for the relative fluctuation we have
\be
\mathcal{R}_{P_S} (t) =  \frac{\langle P_S^2(t) \rangle }{\langle P_S(t) \rangle^2} -1 = \left ( \frac{\langle f_k^2(t) \rangle}{ \langle f_k(t) \rangle^2} \right )^L-1\sim e^{(2\gamma(t)-\gamma_1(t)) L}.
\label{eq:Rrps1}
\ee
Since $2\gamma(t)-\gamma_1(t)>0$, we have that $P_S(t)$ is not self-averaging and $\mathcal{R}_{P_S}$ increases exponentially fast with $L$.

By considering $-\ln{P_S}(t) = -\sum_k \ln{f_k}(t)$ and that the variance of the sum of independent random variables is the sum of the variances of the random variables, 
we obtain that $\sigma_{-\ln{P_S(t)}}^2 = L\sigma_{-\ln{f(t)}}^2$. This leads to the relative variance
\be
\mathcal{R}_{-\ln{P_S}}(t) = \frac{L\sigma_{-\ln{f(t)}}^2}{L^2 \langle-\ln{f(t)} \rangle^2} \propto L^{-1},
\ee
which explains why $-\langle \ln  P_S(t) \rangle$ is self-averaging, as mentioned above.

The fractal dimension associated with the scaling analysis of $\sum |c_{\alpha}^0|^4$ is usually denoted by $\tilde{D}_2$. If it is computed using $-\ln \langle P_S(t> t_{\text{R}}) \rangle $ vs $\ln {\cal D}$, 
the result in Eq.~(\ref{eq:Rrps1}) above implies that $\tilde{D}_2$ is non-self-averaging, but if we use $-\langle \ln  P_S(t> t_{\text{R}}) \rangle $ vs $\ln {\cal D}$ then the fractal dimension is self-averaging.

\subsubsection{Inverse Participation Ratio}

Using the toy model (\ref{eq:H_toy}) for the inverse participation ratio in the many-body localized phase, we get that
\be
 \text{IPR} (t) = \prod_k^L g_k(t),
\ee
where $g_k(t) = |\cos^2 ({\phi_k/2} )e^{-i\epsilon_+^{(k)}} + \sin^2 ({\phi_k/2}) e^{-i\epsilon_-^{(k)}}|^4 + | 2 (\cos{\phi_k/2}) \sin ({\phi_k/2}) \sin( \epsilon_+^{(k)} t)|^4$. 
Using arguments similar to those in the discussion above for the survival probability, it is thus clear that 
$\langle \text{IPR}(t) \rangle \propto e^{-a(t) L}$ and $\mathcal{R}_{\text{IPR}}(t) \propto e^{a_1(t) L}$, with $a(t), a_1(t)\ge 0$.


\subsection{Local Quantities}

We now study the local quantities using the effective Hamiltonian~(\ref{eq:H_eff}) for the $\ell$-bits model.

\subsubsection{Spin Autcorrelation Function} 

\begin{figure}[t]
\includegraphics*[width=0.47\textwidth]{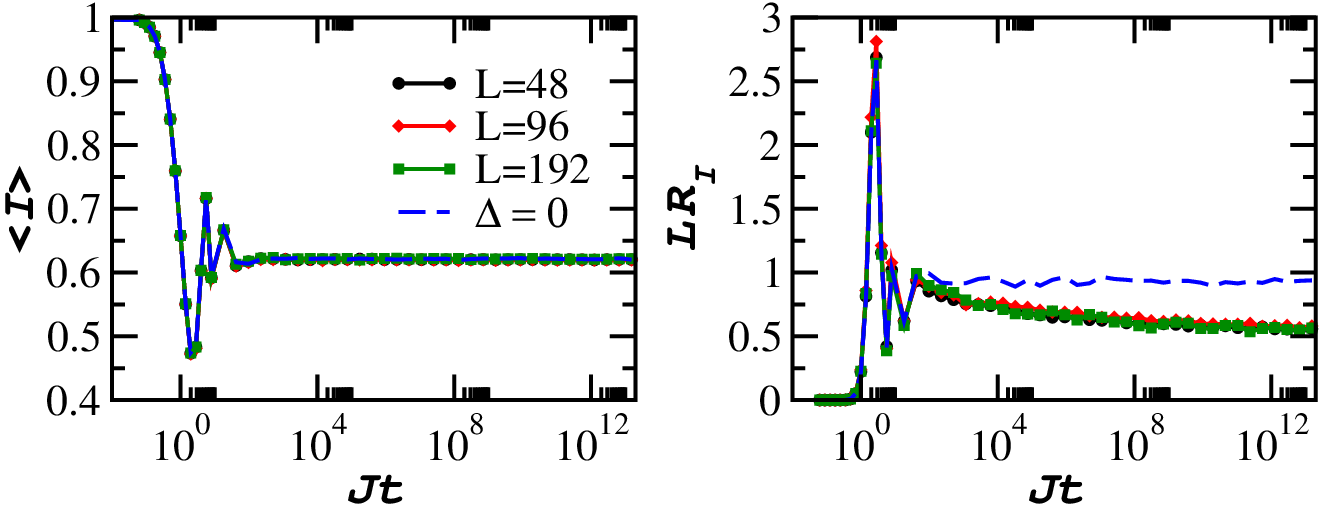}
\caption{Left panels: Mean value of the spin autocorrelation function. Right panels: Relative variance of the spin autocorrelation function. In both panels, the dynamics was computed using the effective $\ell$-bits Hamiltonian $H_{\text{eff}}$ in Eq.~(\ref{eq:H_eff}) with $h=6$ and $\Delta =0.1$ for system sizes $L=48$, $96$, $192$. The dashed-line represents the noninteracting model ($\Delta = 0$) for $L = 48$.  }
\label{fig:Strong_disorder_1}
\end{figure}

Figure~\ref{fig:Strong_disorder_1}~(a) shows $\langle I(t)\rangle$ obtained with $H_{\text{eff}}$~(\ref{eq:H_eff}) for weak interaction, $\Delta  = 0.1$. After a short and quick dynamics, some memory of the initial state is retained and $\langle I(t)\rangle$ saturates to an $L$-independent positive value. This behavior is analogous to what we have for $H$~(\ref{eq:H}) in Fig.~\ref{fig:I}~(k), although there, the saturation point is slightly larger.

The relative variance $\mathcal{R}_I(t)$ rescaled with $L$ is shown in Fig.~\ref{fig:Strong_disorder_1}~(b) making it evident that $\mathcal{R}_I(t)\propto L^{-1}$. The $\ell$-bits model confirms that the spin autcorrelation function is self-averaging at any time in the many-body localized phase, as suggested by the numerical results for  $H$~(\ref{eq:H}) displayed in Fig.~\ref{fig:I}~(l). 

Both panels in Fig.~\ref{fig:Strong_disorder_1} include also the curve for the noninteracting model ($\Delta = 0$). The same scaling, $\mathcal{R}_I(t) \propto L^{-1}$, holds also for this case. The role of the interaction in the $\ell$-bits model is to enhance the self-averaging behavior of $I(t)$ by reducing in time the value of the relative variance $\mathcal{R}_I(t)$ after the interval of oscillations. This contrasts with the noninteracting case, where  $\mathcal{R}_I(t)$ after the oscillations is constant and in fact, more similar to what we see for the available system sizes in Fig.~\ref{fig:I}~(l).

Since the same scaling for $\mathcal{R}_I(t)$ holds for the $\ell$-bits and the noninteracting Hamiltonian, we consider the toy model in Eq.~(\ref{eq:H_toy}) to present an analytical argument supporting the self-averaging property of $I(t)$. One can show that
\begin{equation}
I(t) = \frac{1}{L} \sum_k \{ \cos^2 \phi_k + \sin^2\phi_k \cos[(\epsilon_{+}- \epsilon_{-})t] \},
\end{equation}
where $\sin^2 \phi_k = 1/\sqrt{1+h_k^2}$ and $\epsilon_{\pm} = \pm \sqrt{1+h_k^2}$. The spin autocorrelation function is then a sum of independent identically distributed random variables. Applying the additivity property of the variance, we arrive at $\mathcal{R}_I(t) \propto L^{-1}$.

\subsubsection{Connected Spin-Spin Correlation Function} 
\begin{figure}[t]
\includegraphics*[width=0.47\textwidth]{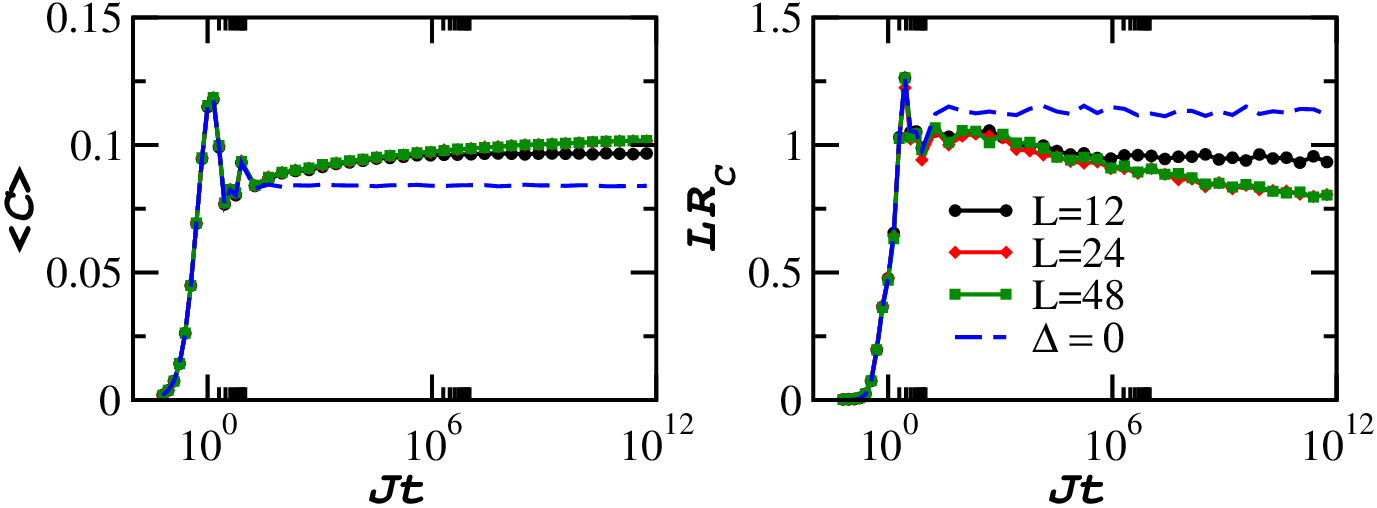}
\caption{Left panels: Mean value of the connected spin-spin correlation function. Right panels: Relative variance of the connected spin autocorrelation function. In both panels, the dynamics was computed using the effective $\ell$-bits Hamiltonian $H_{\text{eff}}$ in Eq.~(\ref{eq:H_eff}) with $h=6$ and $\Delta =0.1$ for system sizes $L=12$, $24$, $48$. The dashed-line represents the noninteracting model ($\Delta = 0$) for $L = 12$. }
\label{fig:Strong_disorder_2}
\end{figure}

Similar conclusions can be drawn for the connected spin-spin correlation function $C(t)$. Figures~\ref{fig:Strong_disorder_2}~(a) and (b) show $|\langle C (t)\rangle |$ and its relative variance $\mathcal{R}_C(t)$ for the $\ell$-bits model. The corresponding results for the noninteracting limit are exhibited with dashed-lines and are similar to those seen in Figs.~\ref{fig:C}~(k) and (l). As in the case of the spin autocorrelation function, we find that $\mathcal{R}_C(t) \propto L^{-1}$ meaning that $\langle C(t) \rangle$ is self-averaging. 

The reason why self-averaging holds for $I(t)$ and $C(t)$  is rooted in the fact that both quantities are sums of expectation values of local observables. At strong disorder, the system can be thought, as a first approximation, as a union of disconnected and independent  small chains and we can thus apply the central limit theorem to understand the behavior of the relative variances with $L$. 

\section{Conclusions}
\label{Sec:Conc}

Based on the analysis of the one-dimensional spin-1/2 Heisenberg model with onsite disorder, this work shows that the self-averaging behavior of many-body quantum systems out of equilibrium is rather non-trivial. It depends on the quantity, time scale, and disorder strength. With the use of the $\ell$-bits model and a toy model, we are able to explain analytically how the mean value and the relative variances of the two non-local and the two local quantities studied here scale with system size in the many-body localized phase. The general picture that we draw from our results is the following. 

(i)  The survival probability, which is non-local in real space and non-local in time, is non-self-averaging at any time scale and for any disorder strength. The behavior is particularly serious in the many-body localized phase, where our analytical results show that the relative variance grows exponentially with system size. These results are worrisome, since $P_S(t)$ is extensively considered in studies of non-equilibrium quantum dynamics and in fundamental questions of quantum mechanics. 

(ii)  The connected spin-spin correlation function measured in experiments with ion traps, which is local in space and in time, is self-averaging at all times and for any disorder strength. In the chaotic region, the relative variance decreases exponentially with system size at long times, while in the many-body localized phase, $\mathcal{R}_C(t) \propto L^{-1}$.

(iii) In between the two above quantities of opposing features, we find the inverse participation ratio, which is non-local in space and local in time, and the spin autocorrelation function, which is local in space and non-local in time. They show complementary behaviors. In the chaotic regime, $\text{IPR}(t)$ is non-self-averaging at short times, but self-averaging at long times, while for $I(t)$, we have just the contrary. As the disorder strength increases, the crossing point between one behavior and the other gets delayed to longer times. Once localization is reached, the inverse participation ratio, just as the survival probability, becomes non-self-averaging at any time scale with a relative variance that grows exponentially with system size, while the spin autocorrelation function, just as the connected spin-spin correlation function, becomes self-averaging at all times with $\mathcal{R}_I(t) \propto L^{-1}$.

(iv) The dependence on the system size of the relative variances of the observables at long times, $t>t_{\text{R}}$, makes a clear distinction between the ergodic and the localized phase. The table below summarizes the differences.

\begin{table}[h]
\begin{tabular}{|c|c|c|}
 \hline
 & chaotic phase & localized phase   \\
 \hline
$\mathcal{R}_{P_S}(t>t_{\text{R}}) $ & $\sim 1$ & $e^{\gamma L}$   \\
  \hline
$\mathcal{R}_{\text{IPR}}(t>t_{\text{R}})$  & $e^{-\gamma L}$ &  $e^{\gamma L}$  \\
  \hline
$\mathcal{R}_{I}(t>t_{\text{R}})$ & grows with $L$ & $1/L$\\  
  \hline
$\mathcal{R}_{C}(t>t_{\text{R}})$ & $e^{-\gamma L}$ & $1/L$ \\  
  \hline
\end{tabular}
\end{table}

Contrary to ergodicity and thermalization, self-averaging in many-body quantum systems out of equilibrium has received very little attention. There are several new questions that can be addressed, from extensions to other isolated time-independent Hamiltonians to time-dependent Hamiltonians and open systems.

\begin{acknowledgments}
E.J.T.-H. acknowledges funding from VIEP-BUAP (Grant Nos. MEBJ-EXC19-G, LUAGEXC19-G), Mexico. He is also grateful to LNS-BUAP for allowing use of their supercomputing facility. M.S. and L.F.S. were supported by the NSF Grant No.~DMR-1603418 and gratefully acknowledges support from the Simons Center for Geometry and Physics, Stony Brook University at which some of the research for this paper was performed.  F.P.B. thanks the Consejer\'ia de Conocimiento, Investigaci\'on y Universidad, Junta de Andaluc\'ia and European Regional Development Fund (ERDF), ref. SOMM17/6105/UGR. Additional computer resources supporting this work were provided by the Universidad de Huelva CEAFMC High Performance Computer  located in the Campus Universitario el Carmen and funded by FEDER/MINECO project UNHU-15CE-2848. L.F.S. is supported by the NSF Grant No.~DMR-1936006. Part of this work was performed at the Aspen Center for Physics, which is supported by National Science Foundation grant PHY-1607611. G.D.T. acknowledges the hospitality of MPIPKS Dresden, where part of the work was performed.
\end{acknowledgments}


\begin{thebibliography}{93}%
\makeatletter
\providecommand \@ifxundefined [1]{%
 \@ifx{#1\undefined}
}%
\providecommand \@ifnum [1]{%
 \ifnum #1\expandafter \@firstoftwo
 \else \expandafter \@secondoftwo
 \fi
}%
\providecommand \@ifx [1]{%
 \ifx #1\expandafter \@firstoftwo
 \else \expandafter \@secondoftwo
 \fi
}%
\providecommand \natexlab [1]{#1}%
\providecommand \enquote  [1]{``#1''}%
\providecommand \bibnamefont  [1]{#1}%
\providecommand \bibfnamefont [1]{#1}%
\providecommand \citenamefont [1]{#1}%
\providecommand \href@noop [0]{\@secondoftwo}%
\providecommand \href [0]{\begingroup \@sanitize@url \@href}%
\providecommand \@href[1]{\@@startlink{#1}\@@href}%
\providecommand \@@href[1]{\endgroup#1\@@endlink}%
\providecommand \@sanitize@url [0]{\catcode `\\12\catcode `\$12\catcode
  `\&12\catcode `\#12\catcode `\^12\catcode `\_12\catcode `\%12\relax}%
\providecommand \@@startlink[1]{}%
\providecommand \@@endlink[0]{}%
\providecommand \url  [0]{\begingroup\@sanitize@url \@url }%
\providecommand \@url [1]{\endgroup\@href {#1}{\urlprefix }}%
\providecommand \urlprefix  [0]{URL }%
\providecommand \Eprint [0]{\href }%
\providecommand \doibase [0]{http://dx.doi.org/}%
\providecommand \selectlanguage [0]{\@gobble}%
\providecommand \bibinfo  [0]{\@secondoftwo}%
\providecommand \bibfield  [0]{\@secondoftwo}%
\providecommand \translation [1]{[#1]}%
\providecommand \BibitemOpen [0]{}%
\providecommand \bibitemStop [0]{}%
\providecommand \bibitemNoStop [0]{.\EOS\space}%
\providecommand \EOS [0]{\spacefactor3000\relax}%
\providecommand \BibitemShut  [1]{\csname bibitem#1\endcsname}%
\let\auto@bib@innerbib\@empty
\bibitem [{\citenamefont {Lifshitz}\ \emph {et~al.}(1988)\citenamefont
  {Lifshitz}, \citenamefont {Gredeskul},\ and\ \citenamefont
  {Pastur}}]{LifshitzBook}%
  \BibitemOpen
  \bibfield  {author} {\bibinfo {author} {\bibfnamefont {I.~M.}\ \bibnamefont
  {Lifshitz}}, \bibinfo {author} {\bibfnamefont {S.~A.}\ \bibnamefont
  {Gredeskul}}, \ and\ \bibinfo {author} {\bibfnamefont {L.~A.}\ \bibnamefont
  {Pastur}},\ }\href@noop {} {\emph {\bibinfo {title} {Introduction to the
  Theory of Disordered Systems}}}\ (\bibinfo  {publisher} {Wiley},\ \bibinfo
  {address} {New York},\ \bibinfo {year} {1988})\BibitemShut {NoStop}%
\bibitem [{\citenamefont {Wiseman}\ and\ \citenamefont
  {Domany}(1995)}]{Wiseman1995}%
  \BibitemOpen
  \bibfield  {author} {\bibinfo {author} {\bibfnamefont {S.}~\bibnamefont
  {Wiseman}}\ and\ \bibinfo {author} {\bibfnamefont {E.}~\bibnamefont
  {Domany}},\ }\bibfield  {title} {\enquote {\bibinfo {title} {Lack of
  self-averaging in critical disordered systems},}\ }\href {\doibase
  10.1103/PhysRevE.52.3469} {\bibfield  {journal} {\bibinfo  {journal} {Phys.
  Rev. E}\ }\textbf {\bibinfo {volume} {52}},\ \bibinfo {pages} {3469}
  (\bibinfo {year} {1995})}\BibitemShut {NoStop}%
\bibitem [{\citenamefont {Aharony}\ and\ \citenamefont
  {Harris}(1996)}]{Aharony1996}%
  \BibitemOpen
  \bibfield  {author} {\bibinfo {author} {\bibfnamefont {A.}~\bibnamefont
  {Aharony}}\ and\ \bibinfo {author} {\bibfnamefont {A.~B.}\ \bibnamefont
  {Harris}},\ }\bibfield  {title} {\enquote {\bibinfo {title} {Absence of
  self-averaging and universal fluctuations in random systems near critical
  points},}\ }\href {\doibase 10.1103/PhysRevLett.77.3700} {\bibfield
  {journal} {\bibinfo  {journal} {Phys. Rev. Lett.}\ }\textbf {\bibinfo
  {volume} {77}},\ \bibinfo {pages} {3700} (\bibinfo {year}
  {1996})}\BibitemShut {NoStop}%
\bibitem [{\citenamefont {Wiseman}\ and\ \citenamefont
  {Domany}(1998)}]{Wiseman1998}%
  \BibitemOpen
  \bibfield  {author} {\bibinfo {author} {\bibfnamefont {S.}~\bibnamefont
  {Wiseman}}\ and\ \bibinfo {author} {\bibfnamefont {E.}~\bibnamefont
  {Domany}},\ }\bibfield  {title} {\enquote {\bibinfo {title} {finite-size
  scaling and lack of self-averaging in critical disordered systems},}\ }\href
  {\doibase 10.1103/PhysRevLett.81.22} {\bibfield  {journal} {\bibinfo
  {journal} {Phys. Rev. Lett.}\ }\textbf {\bibinfo {volume} {81}},\ \bibinfo
  {pages} {22} (\bibinfo {year} {1998})}\BibitemShut {NoStop}%
\bibitem [{\citenamefont {Castellani}\ and\ \citenamefont
  {Cavagna}(2005)}]{Castellani2005}%
  \BibitemOpen
  \bibfield  {author} {\bibinfo {author} {\bibfnamefont {T.}~\bibnamefont
  {Castellani}}\ and\ \bibinfo {author} {\bibfnamefont {A.}~\bibnamefont
  {Cavagna}},\ }\bibfield  {title} {\enquote {\bibinfo {title} {Spin-glass
  theory for pedestrians},}\ }\href {\doibase 10.1088/1742-5468/2005/05/p05012}
  {\bibfield  {journal} {\bibinfo  {journal} {J. Stat. Mech. Th. Exp.}\
  }\textbf {\bibinfo {volume} {2005}},\ \bibinfo {pages} {P05012} (\bibinfo
  {year} {2005})}\BibitemShut {NoStop}%
\bibitem [{\citenamefont {Malakis}\ and\ \citenamefont
  {Fytas}(2006)}]{Malakis2006}%
  \BibitemOpen
  \bibfield  {author} {\bibinfo {author} {\bibfnamefont {A.}~\bibnamefont
  {Malakis}}\ and\ \bibinfo {author} {\bibfnamefont {N.~G.}\ \bibnamefont
  {Fytas}},\ }\bibfield  {title} {\enquote {\bibinfo {title} {Lack of
  self-averaging of the specific heat in the three-dimensional random-field
  {I}sing model},}\ }\href {\doibase 10.1103/PhysRevE.73.016109} {\bibfield
  {journal} {\bibinfo  {journal} {Phys. Rev. E}\ }\textbf {\bibinfo {volume}
  {73}},\ \bibinfo {pages} {016109} (\bibinfo {year} {2006})}\BibitemShut
  {NoStop}%
\bibitem [{\citenamefont {Roy}\ and\ \citenamefont
  {Bhattacharjee}(2006)}]{Roy2006}%
  \BibitemOpen
  \bibfield  {author} {\bibinfo {author} {\bibfnamefont {S.}~\bibnamefont
  {Roy}}\ and\ \bibinfo {author} {\bibfnamefont {S.~M.}\ \bibnamefont
  {Bhattacharjee}},\ }\bibfield  {title} {\enquote {\bibinfo {title} {Is
  small-world network disordered?}}\ }\href {\doibase
  https://doi.org/10.1016/j.physleta.2005.10.105} {\bibfield  {journal}
  {\bibinfo  {journal} {Phys. Lett. A}\ }\textbf {\bibinfo {volume} {352}},\
  \bibinfo {pages} {13} (\bibinfo {year} {2006})}\BibitemShut {NoStop}%
\bibitem [{\citenamefont {Monthus}(2006)}]{Monthus2006}%
  \BibitemOpen
  \bibfield  {author} {\bibinfo {author} {\bibfnamefont {C.}~\bibnamefont
  {Monthus}},\ }\bibfield  {title} {\enquote {\bibinfo {title} {{R}andom
  {W}alks and {P}olymers in the {P}resence of {Q}uenched {D}isorder},}\ }\href
  {\doibase 10.1007/s11005-006-0122-2} {\bibfield  {journal} {\bibinfo
  {journal} {Lett. Math. Phys.}\ }\textbf {\bibinfo {volume} {78}},\ \bibinfo
  {pages} {207} (\bibinfo {year} {2006})}\BibitemShut {NoStop}%
\bibitem [{\citenamefont {Efrat}\ and\ \citenamefont
  {Schwartz}(2014)}]{Efrat2014}%
  \BibitemOpen
  \bibfield  {author} {\bibinfo {author} {\bibfnamefont {A.}~\bibnamefont
  {Efrat}}\ and\ \bibinfo {author} {\bibfnamefont {M.}~\bibnamefont
  {Schwartz}},\ }\bibfield  {title} {\enquote {\bibinfo {title} {{L}ack of
  self-averaging in random systems - {L}iability or asset?}}\ }\href {\doibase
  https://doi.org/10.1016/j.physa.2014.06.071} {\bibfield  {journal} {\bibinfo
  {journal} {Phys. A Stat. Mech. Appl.}\ }\textbf {\bibinfo {volume} {414}},\
  \bibinfo {pages} {137} (\bibinfo {year} {2014})}\BibitemShut {NoStop}%
\bibitem [{\citenamefont {\L{}obejko}\ \emph {et~al.}(2018)\citenamefont
  {\L{}obejko}, \citenamefont {Dajka},\ and\ \citenamefont
  {\L{}uczka}}]{Lobejko2018}%
  \BibitemOpen
  \bibfield  {author} {\bibinfo {author} {\bibfnamefont {M.}~\bibnamefont
  {\L{}obejko}}, \bibinfo {author} {\bibfnamefont {J.}~\bibnamefont {Dajka}}, \
  and\ \bibinfo {author} {\bibfnamefont {J.}~\bibnamefont {\L{}uczka}},\
  }\bibfield  {title} {\enquote {\bibinfo {title} {Self-averaging of random
  quantum dynamics},}\ }\href {\doibase 10.1103/PhysRevA.98.022111} {\bibfield
  {journal} {\bibinfo  {journal} {Phys. Rev. A}\ }\textbf {\bibinfo {volume}
  {98}},\ \bibinfo {pages} {022111} (\bibinfo {year} {2018})}\BibitemShut
  {NoStop}%
\bibitem [{\citenamefont {Bouchaud}\ and\ \citenamefont
  {Georges}(1990)}]{Bouchaud1990}%
  \BibitemOpen
  \bibfield  {author} {\bibinfo {author} {\bibfnamefont {J.-P.}\ \bibnamefont
  {Bouchaud}}\ and\ \bibinfo {author} {\bibfnamefont {A.}~\bibnamefont
  {Georges}},\ }\bibfield  {title} {\enquote {\bibinfo {title} {Anomalous
  diffusion in disordered media: {S}tatistical mechanisms, models and physical
  applications},}\ }\href {\doibase
  https://doi.org/10.1016/0370-1573(90)90099-N} {\bibfield  {journal} {\bibinfo
   {journal} {Phys. Rep.}\ }\textbf {\bibinfo {volume} {195}},\ \bibinfo
  {pages} {127} (\bibinfo {year} {1990})}\BibitemShut {NoStop}%
\bibitem [{\citenamefont {Akimoto}\ \emph {et~al.}(2016)\citenamefont
  {Akimoto}, \citenamefont {Barkai},\ and\ \citenamefont
  {Saito}}]{AkimotoPRL2016}%
  \BibitemOpen
  \bibfield  {author} {\bibinfo {author} {\bibfnamefont {T.}~\bibnamefont
  {Akimoto}}, \bibinfo {author} {\bibfnamefont {E.}~\bibnamefont {Barkai}}, \
  and\ \bibinfo {author} {\bibfnamefont {K.}~\bibnamefont {Saito}},\ }\bibfield
   {title} {\enquote {\bibinfo {title} {{U}niversal {F}luctuations of
  {S}ingle-{P}article {D}iffusivity in a {Q}uenched {E}nvironment},}\ }\href
  {\doibase 10.1103/PhysRevLett.117.180602} {\bibfield  {journal} {\bibinfo
  {journal} {Phys. Rev. Lett.}\ }\textbf {\bibinfo {volume} {117}},\ \bibinfo
  {pages} {180602} (\bibinfo {year} {2016})}\BibitemShut {NoStop}%
\bibitem [{\citenamefont {Russian}\ \emph {et~al.}(2017)\citenamefont
  {Russian}, \citenamefont {Dentz},\ and\ \citenamefont {Gouze}}]{Russian2017}%
  \BibitemOpen
  \bibfield  {author} {\bibinfo {author} {\bibfnamefont {A.}~\bibnamefont
  {Russian}}, \bibinfo {author} {\bibfnamefont {M.}~\bibnamefont {Dentz}}, \
  and\ \bibinfo {author} {\bibfnamefont {P.}~\bibnamefont {Gouze}},\ }\bibfield
   {title} {\enquote {\bibinfo {title} {Self-averaging and weak ergodicity
  breaking of diffusion in heterogeneous media},}\ }\href {\doibase
  10.1103/PhysRevE.96.022156} {\bibfield  {journal} {\bibinfo  {journal} {Phys.
  Rev. E}\ }\textbf {\bibinfo {volume} {96}},\ \bibinfo {pages} {022156}
  (\bibinfo {year} {2017})}\BibitemShut {NoStop}%
\bibitem [{\citenamefont {Akimoto}\ \emph {et~al.}(2018)\citenamefont
  {Akimoto}, \citenamefont {Barkai},\ and\ \citenamefont
  {Saito}}]{AkimotoPRE2018}%
  \BibitemOpen
  \bibfield  {author} {\bibinfo {author} {\bibfnamefont {T.}~\bibnamefont
  {Akimoto}}, \bibinfo {author} {\bibfnamefont {E.}~\bibnamefont {Barkai}}, \
  and\ \bibinfo {author} {\bibfnamefont {K.}~\bibnamefont {Saito}},\ }\bibfield
   {title} {\enquote {\bibinfo {title} {Non-self-averaging behaviors and
  ergodicity in quenched trap models with finite system sizes},}\ }\href
  {\doibase 10.1103/PhysRevE.97.052143} {\bibfield  {journal} {\bibinfo
  {journal} {Phys. Rev. E}\ }\textbf {\bibinfo {volume} {97}},\ \bibinfo
  {pages} {052143} (\bibinfo {year} {2018})}\BibitemShut {NoStop}%
\bibitem [{\citenamefont {Pastur}\ and\ \citenamefont
  {Shcherbina}(1990)}]{Pastur1990}%
  \BibitemOpen
  \bibfield  {author} {\bibinfo {author} {\bibfnamefont {L.}~\bibnamefont
  {Pastur}}\ and\ \bibinfo {author} {\bibfnamefont {M.~V.}\ \bibnamefont
  {Shcherbina}},\ }\bibfield  {title} {\enquote {\bibinfo {title} {Absence of
  self-averaging of the order parameter in the {S}herrington-{K}irkpatrick
  model},}\ }\href {\doibase doi.org/10.1007/BF01020856} {\bibfield  {journal}
  {\bibinfo  {journal} {J. Stat. Phys.}\ }\textbf {\bibinfo {volume} {62}},\
  \bibinfo {pages} {1} (\bibinfo {year} {1990})}\BibitemShut {NoStop}%
\bibitem [{\citenamefont {Wreszinski}\ and\ \citenamefont
  {Bolina}(2004)}]{Wreszinski2004}%
  \BibitemOpen
  \bibfield  {author} {\bibinfo {author} {\bibfnamefont {W.F.}\ \bibnamefont
  {Wreszinski}}\ and\ \bibinfo {author} {\bibfnamefont {O.}~\bibnamefont
  {Bolina}},\ }\bibfield  {title} {\enquote {\bibinfo {title} {A self-averaging
  ``order parameter'' for the {S}herrington-{K}irkpatrick spin glass model},}\
  }\href {\doibase 10.1023/B:JOSS.0000041743.24497.63} {\bibfield  {journal}
  {\bibinfo  {journal} {J. Stat. Phys.}\ }\textbf {\bibinfo {volume} {116}},\
  \bibinfo {pages} {1389} (\bibinfo {year} {2004})}\BibitemShut {NoStop}%
\bibitem [{\citenamefont {Parisi}\ and\ \citenamefont
  {Sourlas}(2002)}]{Parisi2002}%
  \BibitemOpen
  \bibfield  {author} {\bibinfo {author} {\bibfnamefont {Giorgio}\ \bibnamefont
  {Parisi}}\ and\ \bibinfo {author} {\bibfnamefont {Nicolas}\ \bibnamefont
  {Sourlas}},\ }\bibfield  {title} {\enquote {\bibinfo {title} {Scale
  invariance in disordered systems: The example of the random-field {I}sing
  model},}\ }\href {\doibase 10.1103/PhysRevLett.89.257204} {\bibfield
  {journal} {\bibinfo  {journal} {Phys. Rev. Lett.}\ }\textbf {\bibinfo
  {volume} {89}},\ \bibinfo {pages} {257204} (\bibinfo {year}
  {2002})}\BibitemShut {NoStop}%
\bibitem [{\citenamefont {M\"uller}\ and\ \citenamefont
  {Delande}()}]{MullerARXIV}%
  \BibitemOpen
  \bibfield  {author} {\bibinfo {author} {\bibfnamefont {C.~A.}\ \bibnamefont
  {M\"uller}}\ and\ \bibinfo {author} {\bibfnamefont {D.}~\bibnamefont
  {Delande}},\ }\href@noop {} {\enquote {\bibinfo {title} {{D}isorder and
  interference: localization phenomena},}\ }\bibinfo {note} {Les Houches 2009 -
  Session XCI: Ultracold Gases and Quantum Information, C. Miniatura, L.-C.
  Kwek, M. Ducloy, B. Gremaud, B.-G. Englert, L.F. Cugliandolo, A. Ekert, eds.
  (Oxford University Press, Oxford 2011); arXiv:1004.0915}\BibitemShut
  {NoStop}%
\bibitem [{\citenamefont {Pastur}\ and\ \citenamefont
  {Slavin}(2014)}]{Pastur2014PRL}%
  \BibitemOpen
  \bibfield  {author} {\bibinfo {author} {\bibfnamefont {L.}~\bibnamefont
  {Pastur}}\ and\ \bibinfo {author} {\bibfnamefont {V.}~\bibnamefont
  {Slavin}},\ }\bibfield  {title} {\enquote {\bibinfo {title} {{A}rea {L}aw
  {S}caling for the {E}ntropy of {D}isordered {Q}uasifree {F}ermions},}\ }\href
  {\doibase 10.1103/PhysRevLett.113.150404} {\bibfield  {journal} {\bibinfo
  {journal} {Phys. Rev. Lett.}\ }\textbf {\bibinfo {volume} {113}},\ \bibinfo
  {pages} {150404} (\bibinfo {year} {2014})}\BibitemShut {NoStop}%
\bibitem [{\citenamefont {Santos}\ \emph {et~al.}(2004)\citenamefont {Santos},
  \citenamefont {Rigolin},\ and\ \citenamefont {Escobar}}]{SantosEscobar2004}%
  \BibitemOpen
  \bibfield  {author} {\bibinfo {author} {\bibfnamefont {L.~F.}\ \bibnamefont
  {Santos}}, \bibinfo {author} {\bibfnamefont {G.}~\bibnamefont {Rigolin}}, \
  and\ \bibinfo {author} {\bibfnamefont {C.~O.}\ \bibnamefont {Escobar}},\
  }\bibfield  {title} {\enquote {\bibinfo {title} {Entanglement versus chaos in
  disordered spin systems},}\ }\href@noop {} {\bibfield  {journal} {\bibinfo
  {journal} {Phys. Rev. A}\ }\textbf {\bibinfo {volume} {69}},\ \bibinfo
  {pages} {042304} (\bibinfo {year} {2004})}\BibitemShut {NoStop}%
\bibitem [{\citenamefont {Basko}\ \emph {et~al.}(2006)\citenamefont {Basko},
  \citenamefont {Aleiner},\ and\ \citenamefont {Altshuler}}]{Basko2006}%
  \BibitemOpen
  \bibfield  {author} {\bibinfo {author} {\bibfnamefont {D.~M.}\ \bibnamefont
  {Basko}}, \bibinfo {author} {\bibfnamefont {I.~L.}\ \bibnamefont {Aleiner}},
  \ and\ \bibinfo {author} {\bibfnamefont {B.~L.}\ \bibnamefont {Altshuler}},\
  }\bibfield  {title} {\enquote {\bibinfo {title} {Metal-insulator transition
  in a weakly interacting many-electron system with localized single-particle
  states},}\ }\href@noop {} {\bibfield  {journal} {\bibinfo  {journal} {Ann.
  Phys.}\ }\textbf {\bibinfo {volume} {321}},\ \bibinfo {pages} {1126}
  (\bibinfo {year} {2006})}\BibitemShut {NoStop}%
\bibitem [{\citenamefont {Nandkishore}\ and\ \citenamefont
  {Huse}(2015)}]{Nandkishore2015}%
  \BibitemOpen
  \bibfield  {author} {\bibinfo {author} {\bibfnamefont {R.}~\bibnamefont
  {Nandkishore}}\ and\ \bibinfo {author} {\bibfnamefont {D.A.}\ \bibnamefont
  {Huse}},\ }\bibfield  {title} {\enquote {\bibinfo {title} {Many-body
  localization and thermalization in quantum statistical mechanics},}\
  }\href@noop {} {\bibfield  {journal} {\bibinfo  {journal} {Annu. Rev.
  Condens. Matter Phys.}\ }\textbf {\bibinfo {volume} {6}},\ \bibinfo {pages}
  {15} (\bibinfo {year} {2015})}\BibitemShut {NoStop}%
\bibitem [{\citenamefont {Luitz}\ and\ \citenamefont {Lev}(2017)}]{Luitz2017}%
  \BibitemOpen
  \bibfield  {author} {\bibinfo {author} {\bibfnamefont {D.}~\bibnamefont
  {Luitz}}\ and\ \bibinfo {author} {\bibfnamefont {Y.~Bar}\ \bibnamefont
  {Lev}},\ }\bibfield  {title} {\enquote {\bibinfo {title} {The ergodic side of
  the many-body localization transition},}\ }\href {\doibase
  10.1002/andp.201600350} {\bibfield  {journal} {\bibinfo  {journal} {Ann.
  Phys.(Berlin)}\ }\textbf {\bibinfo {volume} {529}},\ \bibinfo {pages}
  {1600350} (\bibinfo {year} {2017})}\BibitemShut {NoStop}%
\bibitem [{\citenamefont {Altman}(2018)}]{Altman2018}%
  \BibitemOpen
  \bibfield  {author} {\bibinfo {author} {\bibfnamefont {Ehud}\ \bibnamefont
  {Altman}},\ }\bibfield  {title} {\enquote {\bibinfo {title} {Many-body
  localization and quantum thermalization},}\ }\href {\doibase
  10.1038/s41567-018-0305-7} {\bibfield  {journal} {\bibinfo  {journal} {Nat.
  Phys.}\ }\textbf {\bibinfo {volume} {14}},\ \bibinfo {pages} {979--983}
  (\bibinfo {year} {2018})}\BibitemShut {NoStop}%
\bibitem [{\citenamefont {Doggen}\ \emph {et~al.}(2018)\citenamefont {Doggen},
  \citenamefont {Schindler}, \citenamefont {Tikhonov}, \citenamefont {Mirlin},
  \citenamefont {Neupert}, \citenamefont {Polyakov},\ and\ \citenamefont
  {Gornyi}}]{Doggen2018}%
  \BibitemOpen
  \bibfield  {author} {\bibinfo {author} {\bibfnamefont {Elmer V.~H.}\
  \bibnamefont {Doggen}}, \bibinfo {author} {\bibfnamefont {Frank}\
  \bibnamefont {Schindler}}, \bibinfo {author} {\bibfnamefont {Konstantin~S.}\
  \bibnamefont {Tikhonov}}, \bibinfo {author} {\bibfnamefont {Alexander~D.}\
  \bibnamefont {Mirlin}}, \bibinfo {author} {\bibfnamefont {Titus}\
  \bibnamefont {Neupert}}, \bibinfo {author} {\bibfnamefont {Dmitry~G.}\
  \bibnamefont {Polyakov}}, \ and\ \bibinfo {author} {\bibfnamefont {Igor~V.}\
  \bibnamefont {Gornyi}},\ }\bibfield  {title} {\enquote {\bibinfo {title}
  {Many-body localization and delocalization in large quantum chains},}\ }\href
  {\doibase 10.1103/PhysRevB.98.174202} {\bibfield  {journal} {\bibinfo
  {journal} {Phys. Rev. B}\ }\textbf {\bibinfo {volume} {98}},\ \bibinfo
  {pages} {174202} (\bibinfo {year} {2018})}\BibitemShut {NoStop}%
\bibitem [{\citenamefont {Bari\ifmmode \check{s}\else
  \v{s}\fi{}i\ifmmode~\acute{c}\else \'{c}\fi{}}\ \emph
  {et~al.}(2016)\citenamefont {Bari\ifmmode \check{s}\else
  \v{s}\fi{}i\ifmmode~\acute{c}\else \'{c}\fi{}}, \citenamefont {Kokalj},
  \citenamefont {Balog},\ and\ \citenamefont {Prelov\ifmmode~\check{s}\else
  \v{s}\fi{}ek}}]{Barisic2016}%
  \BibitemOpen
  \bibfield  {author} {\bibinfo {author} {\bibfnamefont {Osor~S.}\ \bibnamefont
  {Bari\ifmmode \check{s}\else \v{s}\fi{}i\ifmmode~\acute{c}\else \'{c}\fi{}}},
  \bibinfo {author} {\bibfnamefont {Jure}\ \bibnamefont {Kokalj}}, \bibinfo
  {author} {\bibfnamefont {Ivan}\ \bibnamefont {Balog}}, \ and\ \bibinfo
  {author} {\bibfnamefont {Peter}\ \bibnamefont {Prelov\ifmmode~\check{s}\else
  \v{s}\fi{}ek}},\ }\bibfield  {title} {\enquote {\bibinfo {title} {Dynamical
  conductivity and its fluctuations along the crossover to many-body
  localization},}\ }\href {\doibase 10.1103/PhysRevB.94.045126} {\bibfield
  {journal} {\bibinfo  {journal} {Phys. Rev. B}\ }\textbf {\bibinfo {volume}
  {94}},\ \bibinfo {pages} {045126} (\bibinfo {year} {2016})}\BibitemShut
  {NoStop}%
\bibitem [{\citenamefont {Prelov\ifmmode~\check{s}\else \v{s}\fi{}ek}\ \emph
  {et~al.}(2017)\citenamefont {Prelov\ifmmode~\check{s}\else \v{s}\fi{}ek},
  \citenamefont {Mierzejewski}, \citenamefont {Bari\ifmmode \check{s}\else
  \v{s}\fi{}i\ifmmode~\acute{c}\else \'{c}\fi{}},\ and\ \citenamefont
  {Herbrych}}]{Prelovsek2017}%
  \BibitemOpen
  \bibfield  {author} {\bibinfo {author} {\bibfnamefont {P.}~\bibnamefont
  {Prelov\ifmmode~\check{s}\else \v{s}\fi{}ek}}, \bibinfo {author}
  {\bibfnamefont {M.}~\bibnamefont {Mierzejewski}}, \bibinfo {author}
  {\bibfnamefont {O.}~\bibnamefont {Bari\ifmmode \check{s}\else
  \v{s}\fi{}i\ifmmode~\acute{c}\else \'{c}\fi{}}}, \ and\ \bibinfo {author}
  {\bibfnamefont {J.}~\bibnamefont {Herbrych}},\ }\bibfield  {title} {\enquote
  {\bibinfo {title} {Density correlations and transport in models of many-body
  localization},}\ }\href {\doibase 10.1002/andp.201600362} {\bibfield
  {journal} {\bibinfo  {journal} {Ann. Phys. (Berlin)}\ }\textbf {\bibinfo
  {volume} {529}},\ \bibinfo {pages} {1600362} (\bibinfo {year}
  {2017})}\BibitemShut {NoStop}%
\bibitem [{\citenamefont {Khemani}\ \emph {et~al.}(2017)\citenamefont
  {Khemani}, \citenamefont {Lim}, \citenamefont {Sheng},\ and\ \citenamefont
  {Huse}}]{Khemani2017}%
  \BibitemOpen
  \bibfield  {author} {\bibinfo {author} {\bibfnamefont {Vedika}\ \bibnamefont
  {Khemani}}, \bibinfo {author} {\bibfnamefont {S.{\hspace{0.167em}}P.}\
  \bibnamefont {Lim}}, \bibinfo {author} {\bibfnamefont
  {D.{\hspace{0.167em}}N.}\ \bibnamefont {Sheng}}, \ and\ \bibinfo {author}
  {\bibfnamefont {David~A.}\ \bibnamefont {Huse}},\ }\bibfield  {title}
  {\enquote {\bibinfo {title} {Critical properties of the many-body
  localization transition},}\ }\href {\doibase 10.1103/physrevx.7.021013}
  {\bibfield  {journal} {\bibinfo  {journal} {Physical Review X}\ }\textbf
  {\bibinfo {volume} {7}} (\bibinfo {year} {2017}),\
  10.1103/physrevx.7.021013}\BibitemShut {NoStop}%
\bibitem [{\citenamefont {Serbyn}\ \emph {et~al.}(2017)\citenamefont {Serbyn},
  \citenamefont {Papi\ifmmode~\acute{c}\else \'{c}\fi{}},\ and\ \citenamefont
  {Abanin}}]{Serbyn2017}%
  \BibitemOpen
  \bibfield  {author} {\bibinfo {author} {\bibfnamefont {M.}~\bibnamefont
  {Serbyn}}, \bibinfo {author} {\bibfnamefont {Z.}~\bibnamefont
  {Papi\ifmmode~\acute{c}\else \'{c}\fi{}}}, \ and\ \bibinfo {author}
  {\bibfnamefont {D.~A.}\ \bibnamefont {Abanin}},\ }\bibfield  {title}
  {\enquote {\bibinfo {title} {Thouless energy and multifractality across the
  many-body localization transition},}\ }\href {\doibase
  10.1103/PhysRevB.96.104201} {\bibfield  {journal} {\bibinfo  {journal} {Phys.
  Rev. B}\ }\textbf {\bibinfo {volume} {96}},\ \bibinfo {pages} {104201}
  (\bibinfo {year} {2017})}\BibitemShut {NoStop}%
\bibitem [{\citenamefont {Richter}\ \emph {et~al.}(2020)\citenamefont
  {Richter}, \citenamefont {Schubert},\ and\ \citenamefont
  {Steinigeweg}}]{Richter2020}%
  \BibitemOpen
  \bibfield  {author} {\bibinfo {author} {\bibfnamefont {Jonas}\ \bibnamefont
  {Richter}}, \bibinfo {author} {\bibfnamefont {Dennis}\ \bibnamefont
  {Schubert}}, \ and\ \bibinfo {author} {\bibfnamefont {Robin}\ \bibnamefont
  {Steinigeweg}},\ }\bibfield  {title} {\enquote {\bibinfo {title} {Decay of
  spin-spin correlations in disordered quantum and classical spin chains},}\
  }\href {\doibase 10.1103/PhysRevResearch.2.013130} {\bibfield  {journal}
  {\bibinfo  {journal} {Phys. Rev. Research}\ }\textbf {\bibinfo {volume}
  {2}},\ \bibinfo {pages} {013130} (\bibinfo {year} {2020})}\BibitemShut
  {NoStop}%
\bibitem [{\citenamefont {Ithier}\ and\ \citenamefont
  {Benaych-Georges}(2017)}]{Ithier2017}%
  \BibitemOpen
  \bibfield  {author} {\bibinfo {author} {\bibfnamefont {Gr{\'{e}}goire}\
  \bibnamefont {Ithier}}\ and\ \bibinfo {author} {\bibfnamefont {Florent}\
  \bibnamefont {Benaych-Georges}},\ }\bibfield  {title} {\enquote {\bibinfo
  {title} {Dynamical typicality of embedded quantum systems},}\ }\href
  {\doibase 10.1103/physreva.96.012108} {\bibfield  {journal} {\bibinfo
  {journal} {Phys. Rev. A}\ }\textbf {\bibinfo {volume} {96}} (\bibinfo {year}
  {2017}),\ 10.1103/physreva.96.012108}\BibitemShut {NoStop}%
\bibitem [{\citenamefont {Mukherjee}(2018)}]{Mukherjee2018}%
  \BibitemOpen
  \bibfield  {author} {\bibinfo {author} {\bibfnamefont {B.}~\bibnamefont
  {Mukherjee}},\ }\bibfield  {title} {\enquote {\bibinfo {title} {Floquet
  topological transition by unpolarized light},}\ }\href {\doibase
  10.1103/PhysRevB.98.235112} {\bibfield  {journal} {\bibinfo  {journal} {Phys.
  Rev. B}\ }\textbf {\bibinfo {volume} {98}},\ \bibinfo {pages} {235112}
  (\bibinfo {year} {2018})}\BibitemShut {NoStop}%
\bibitem [{\citenamefont {Argaman}\ \emph {et~al.}(1993)\citenamefont
  {Argaman}, \citenamefont {Dittes}, \citenamefont {Doron}, \citenamefont
  {Keating}, \citenamefont {Kitaev}, \citenamefont {Sieber},\ and\
  \citenamefont {Smilansky}}]{Argaman1993b}%
  \BibitemOpen
  \bibfield  {author} {\bibinfo {author} {\bibfnamefont {N.}~\bibnamefont
  {Argaman}}, \bibinfo {author} {\bibfnamefont {F.-M.}\ \bibnamefont {Dittes}},
  \bibinfo {author} {\bibfnamefont {E.}~\bibnamefont {Doron}}, \bibinfo
  {author} {\bibfnamefont {J.~P.}\ \bibnamefont {Keating}}, \bibinfo {author}
  {\bibfnamefont {A.~Yu.}\ \bibnamefont {Kitaev}}, \bibinfo {author}
  {\bibfnamefont {M.}~\bibnamefont {Sieber}}, \ and\ \bibinfo {author}
  {\bibfnamefont {U.}~\bibnamefont {Smilansky}},\ }\bibfield  {title} {\enquote
  {\bibinfo {title} {Correlations in the actions of periodic orbits derived
  from quantum chaos},}\ }\href {\doibase 10.1103/PhysRevLett.71.4326}
  {\bibfield  {journal} {\bibinfo  {journal} {Phys. Rev. Lett.}\ }\textbf
  {\bibinfo {volume} {71}},\ \bibinfo {pages} {4326--4329} (\bibinfo {year}
  {1993})}\BibitemShut {NoStop}%
\bibitem [{\citenamefont {Eckhardt}\ and\ \citenamefont
  {Main}(1995)}]{Eckhardt1995}%
  \BibitemOpen
  \bibfield  {author} {\bibinfo {author} {\bibfnamefont {B.}~\bibnamefont
  {Eckhardt}}\ and\ \bibinfo {author} {\bibfnamefont {J.}~\bibnamefont
  {Main}},\ }\bibfield  {title} {\enquote {\bibinfo {title} {{S}emiclassical
  {F}orm {F}actor of {M}atrix {E}lement {F}luctuations},}\ }\href {\doibase
  10.1103/PhysRevLett.75.2300} {\bibfield  {journal} {\bibinfo  {journal}
  {Phys. Rev. Lett.}\ }\textbf {\bibinfo {volume} {75}},\ \bibinfo {pages}
  {2300--2303} (\bibinfo {year} {1995})}\BibitemShut {NoStop}%
\bibitem [{\citenamefont {Prange}(1997)}]{Prange1997}%
  \BibitemOpen
  \bibfield  {author} {\bibinfo {author} {\bibfnamefont {R.~E.}\ \bibnamefont
  {Prange}},\ }\bibfield  {title} {\enquote {\bibinfo {title} {The spectral
  form factor is not self-averaging},}\ }\href {\doibase
  10.1103/PhysRevLett.78.2280} {\bibfield  {journal} {\bibinfo  {journal}
  {Phys. Rev. Lett.}\ }\textbf {\bibinfo {volume} {78}},\ \bibinfo {pages}
  {2280} (\bibinfo {year} {1997})}\BibitemShut {NoStop}%
\bibitem [{\citenamefont {Braun}\ and\ \citenamefont
  {Haake}(2015)}]{Braun2015}%
  \BibitemOpen
  \bibfield  {author} {\bibinfo {author} {\bibfnamefont {Petr}\ \bibnamefont
  {Braun}}\ and\ \bibinfo {author} {\bibfnamefont {Fritz}\ \bibnamefont
  {Haake}},\ }\bibfield  {title} {\enquote {\bibinfo {title} {Self-averaging
  characteristics of spectral fluctuations},}\ }\href {\doibase
  10.1088/1751-8113/48/13/135101} {\bibfield  {journal} {\bibinfo  {journal}
  {J.Phys. A}\ }\textbf {\bibinfo {volume} {48}},\ \bibinfo {pages} {135101}
  (\bibinfo {year} {2015})}\BibitemShut {NoStop}%
\bibitem [{\citenamefont {Schiulaz}\ \emph {et~al.}(2020)\citenamefont
  {Schiulaz}, \citenamefont {Torres-Herrera}, \citenamefont {P\'erez-Bernal},\
  and\ \citenamefont {Santos}}]{Schiulaz2020}%
  \BibitemOpen
  \bibfield  {author} {\bibinfo {author} {\bibfnamefont {Mauro}\ \bibnamefont
  {Schiulaz}}, \bibinfo {author} {\bibfnamefont {E.~Jonathan}\ \bibnamefont
  {Torres-Herrera}}, \bibinfo {author} {\bibfnamefont {Francisco}\ \bibnamefont
  {P\'erez-Bernal}}, \ and\ \bibinfo {author} {\bibfnamefont {Lea~F.}\
  \bibnamefont {Santos}},\ }\bibfield  {title} {\enquote {\bibinfo {title}
  {Self-averaging in many-body quantum systems out of equilibrium: Chaotic
  systems},}\ }\href {\doibase 10.1103/PhysRevB.101.174312} {\bibfield
  {journal} {\bibinfo  {journal} {Phys. Rev. B}\ }\textbf {\bibinfo {volume}
  {101}},\ \bibinfo {pages} {174312} (\bibinfo {year} {2020})}\BibitemShut
  {NoStop}%
\bibitem [{\citenamefont {Schreiber}\ \emph {et~al.}(2015)\citenamefont
  {Schreiber}, \citenamefont {Hodgman}, \citenamefont {Bordia}, \citenamefont
  {L{\"u}schen}, \citenamefont {Fischer}, \citenamefont {Vosk}, \citenamefont
  {Altman}, \citenamefont {Schneider},\ and\ \citenamefont
  {Bloch}}]{Schreiber2015}%
  \BibitemOpen
  \bibfield  {author} {\bibinfo {author} {\bibfnamefont {M.}~\bibnamefont
  {Schreiber}}, \bibinfo {author} {\bibfnamefont {S.~S.}\ \bibnamefont
  {Hodgman}}, \bibinfo {author} {\bibfnamefont {Pr.}\ \bibnamefont {Bordia}},
  \bibinfo {author} {\bibfnamefont {H.~P.}\ \bibnamefont {L{\"u}schen}},
  \bibinfo {author} {\bibfnamefont {M.~H.}\ \bibnamefont {Fischer}}, \bibinfo
  {author} {\bibfnamefont {R.}~\bibnamefont {Vosk}}, \bibinfo {author}
  {\bibfnamefont {E.}~\bibnamefont {Altman}}, \bibinfo {author} {\bibfnamefont
  {U.}~\bibnamefont {Schneider}}, \ and\ \bibinfo {author} {\bibfnamefont
  {I.}~\bibnamefont {Bloch}},\ }\bibfield  {title} {\enquote {\bibinfo {title}
  {Observation of many-body localization of interacting fermions in a
  quasirandom optical lattice},}\ }\href {\doibase 10.1126/science.aaa7432}
  {\bibfield  {journal} {\bibinfo  {journal} {Science}\ }\textbf {\bibinfo
  {volume} {349}},\ \bibinfo {pages} {842--845} (\bibinfo {year}
  {2015})}\BibitemShut {NoStop}%
\bibitem [{\citenamefont {Richerme}\ \emph {et~al.}(2014)\citenamefont
  {Richerme}, \citenamefont {Gong}, \citenamefont {Lee}, \citenamefont {Senko},
  \citenamefont {Smith}, \citenamefont {Foss-Feig}, \citenamefont {Michalakis},
  \citenamefont {Gorshkov},\ and\ \citenamefont {Monroe}}]{Richerme2014}%
  \BibitemOpen
  \bibfield  {author} {\bibinfo {author} {\bibfnamefont {P.}~\bibnamefont
  {Richerme}}, \bibinfo {author} {\bibfnamefont {Z.-X.}\ \bibnamefont {Gong}},
  \bibinfo {author} {\bibfnamefont {A.}~\bibnamefont {Lee}}, \bibinfo {author}
  {\bibfnamefont {Cr.}\ \bibnamefont {Senko}}, \bibinfo {author} {\bibfnamefont
  {J.}~\bibnamefont {Smith}}, \bibinfo {author} {\bibfnamefont
  {M.}~\bibnamefont {Foss-Feig}}, \bibinfo {author} {\bibfnamefont
  {S.}~\bibnamefont {Michalakis}}, \bibinfo {author} {\bibfnamefont {A.~V.}\
  \bibnamefont {Gorshkov}}, \ and\ \bibinfo {author} {\bibfnamefont
  {C.}~\bibnamefont {Monroe}},\ }\bibfield  {title} {\enquote {\bibinfo {title}
  {Non-local propagation of correlations in quantum systems with long-range
  interactions},}\ }\href@noop {} {\bibfield  {journal} {\bibinfo  {journal}
  {Nature}\ }\textbf {\bibinfo {volume} {511}},\ \bibinfo {pages} {198--201}
  (\bibinfo {year} {2014})}\BibitemShut {NoStop}%
\bibitem [{\citenamefont {Serbyn}\ \emph
  {et~al.}(2013{\natexlab{a}})\citenamefont {Serbyn}, \citenamefont
  {Papi\ifmmode~\acute{c}\else \'{c}\fi{}},\ and\ \citenamefont
  {Abanin}}]{Serbyn2013}%
  \BibitemOpen
  \bibfield  {author} {\bibinfo {author} {\bibfnamefont {Maksym}\ \bibnamefont
  {Serbyn}}, \bibinfo {author} {\bibfnamefont {Z.}~\bibnamefont
  {Papi\ifmmode~\acute{c}\else \'{c}\fi{}}}, \ and\ \bibinfo {author}
  {\bibfnamefont {Dmitry~A.}\ \bibnamefont {Abanin}},\ }\bibfield  {title}
  {\enquote {\bibinfo {title} {Local conservation laws and the structure of the
  many-body localized states},}\ }\href {\doibase
  10.1103/PhysRevLett.111.127201} {\bibfield  {journal} {\bibinfo  {journal}
  {Phys. Rev. Lett.}\ }\textbf {\bibinfo {volume} {111}},\ \bibinfo {pages}
  {127201} (\bibinfo {year} {2013}{\natexlab{a}})}\BibitemShut {NoStop}%
\bibitem [{\citenamefont {Huse}\ \emph {et~al.}(2014)\citenamefont {Huse},
  \citenamefont {Nandkishore},\ and\ \citenamefont {Oganesyan}}]{Huse2014}%
  \BibitemOpen
  \bibfield  {author} {\bibinfo {author} {\bibfnamefont {David~A.}\
  \bibnamefont {Huse}}, \bibinfo {author} {\bibfnamefont {Rahul}\ \bibnamefont
  {Nandkishore}}, \ and\ \bibinfo {author} {\bibfnamefont {Vadim}\ \bibnamefont
  {Oganesyan}},\ }\bibfield  {title} {\enquote {\bibinfo {title} {Phenomenology
  of fully many-body-localized systems},}\ }\href {\doibase
  10.1103/PhysRevB.90.174202} {\bibfield  {journal} {\bibinfo  {journal} {Phys.
  Rev. B}\ }\textbf {\bibinfo {volume} {90}},\ \bibinfo {pages} {174202}
  (\bibinfo {year} {2014})}\BibitemShut {NoStop}%
\bibitem [{\citenamefont {Iemini}\ \emph {et~al.}(2016)\citenamefont {Iemini},
  \citenamefont {Russomanno}, \citenamefont {Rossini}, \citenamefont
  {Scardicchio},\ and\ \citenamefont {Fazio}}]{Iemini2016}%
  \BibitemOpen
  \bibfield  {author} {\bibinfo {author} {\bibfnamefont {Fernando}\
  \bibnamefont {Iemini}}, \bibinfo {author} {\bibfnamefont {Angelo}\
  \bibnamefont {Russomanno}}, \bibinfo {author} {\bibfnamefont {Davide}\
  \bibnamefont {Rossini}}, \bibinfo {author} {\bibfnamefont {Antonello}\
  \bibnamefont {Scardicchio}}, \ and\ \bibinfo {author} {\bibfnamefont
  {Rosario}\ \bibnamefont {Fazio}},\ }\bibfield  {title} {\enquote {\bibinfo
  {title} {Signatures of many-body localization in the dynamics of two-site
  entanglement},}\ }\href {\doibase 10.1103/PhysRevB.94.214206} {\bibfield
  {journal} {\bibinfo  {journal} {Phys. Rev. B}\ }\textbf {\bibinfo {volume}
  {94}},\ \bibinfo {pages} {214206} (\bibinfo {year} {2016})}\BibitemShut
  {NoStop}%
\bibitem [{\citenamefont {De~Tomasi}\ \emph {et~al.}(2019)\citenamefont
  {De~Tomasi}, \citenamefont {Pollmann},\ and\ \citenamefont
  {Heyl}}]{Tomasi2019}%
  \BibitemOpen
  \bibfield  {author} {\bibinfo {author} {\bibfnamefont {Giuseppe}\
  \bibnamefont {De~Tomasi}}, \bibinfo {author} {\bibfnamefont {Frank}\
  \bibnamefont {Pollmann}}, \ and\ \bibinfo {author} {\bibfnamefont {Markus}\
  \bibnamefont {Heyl}},\ }\bibfield  {title} {\enquote {\bibinfo {title}
  {Efficiently solving the dynamics of many-body localized systems at strong
  disorder},}\ }\href {\doibase 10.1103/PhysRevB.99.241114} {\bibfield
  {journal} {\bibinfo  {journal} {Phys. Rev. B}\ }\textbf {\bibinfo {volume}
  {99}},\ \bibinfo {pages} {241114} (\bibinfo {year} {2019})}\BibitemShut
  {NoStop}%
\bibitem [{\citenamefont {Srednicki}(1996)}]{Srednicki1996}%
  \BibitemOpen
  \bibfield  {author} {\bibinfo {author} {\bibfnamefont {M.}~\bibnamefont
  {Srednicki}},\ }\bibfield  {title} {\enquote {\bibinfo {title} {Thermal
  fluctuations in quantized chaotic systems},}\ }\href@noop {} {\bibfield
  {journal} {\bibinfo  {journal} {J. Phys. A}\ }\textbf {\bibinfo {volume}
  {29}},\ \bibinfo {pages} {L75--L79} (\bibinfo {year} {1996})}\BibitemShut
  {NoStop}%
\bibitem [{\citenamefont {Reimann}(2008)}]{Reimann2008}%
  \BibitemOpen
  \bibfield  {author} {\bibinfo {author} {\bibfnamefont {Peter}\ \bibnamefont
  {Reimann}},\ }\bibfield  {title} {\enquote {\bibinfo {title} {Foundation of
  statistical mechanics under experimentally realistic conditions},}\
  }\href@noop {} {\bibfield  {journal} {\bibinfo  {journal} {Phys. Rev. Lett.}\
  }\textbf {\bibinfo {volume} {101}},\ \bibinfo {pages} {190403} (\bibinfo
  {year} {2008})}\BibitemShut {NoStop}%
\bibitem [{\citenamefont {Short}(2011)}]{Short2011}%
  \BibitemOpen
  \bibfield  {author} {\bibinfo {author} {\bibfnamefont {A.~J.}\ \bibnamefont
  {Short}},\ }\bibfield  {title} {\enquote {\bibinfo {title} {Equilibration of
  quantum systems and subsystems},}\ }\href@noop {} {\bibfield  {journal}
  {\bibinfo  {journal} {New J. Phys.}\ }\textbf {\bibinfo {volume} {13}},\
  \bibinfo {pages} {053009} (\bibinfo {year} {2011})}\BibitemShut {NoStop}%
\bibitem [{\citenamefont {Zangara}\ \emph {et~al.}(2013)\citenamefont
  {Zangara}, \citenamefont {Dente}, \citenamefont {Torres-Herrera},
  \citenamefont {Pastawski}, \citenamefont {Iucci},\ and\ \citenamefont
  {Santos}}]{Zangara2013}%
  \BibitemOpen
  \bibfield  {author} {\bibinfo {author} {\bibfnamefont {Pablo~R.}\
  \bibnamefont {Zangara}}, \bibinfo {author} {\bibfnamefont {Axel~D.}\
  \bibnamefont {Dente}}, \bibinfo {author} {\bibfnamefont {E.~J.}\ \bibnamefont
  {Torres-Herrera}}, \bibinfo {author} {\bibfnamefont {Horacio~M.}\
  \bibnamefont {Pastawski}}, \bibinfo {author} {\bibfnamefont {A.}~\bibnamefont
  {Iucci}}, \ and\ \bibinfo {author} {\bibfnamefont {Lea~F.}\ \bibnamefont
  {Santos}},\ }\bibfield  {title} {\enquote {\bibinfo {title} {Time
  fluctuations in isolated quantum systems of interacting particles},}\
  }\href@noop {} {\bibfield  {journal} {\bibinfo  {journal} {Phys. Rev. E}\
  }\textbf {\bibinfo {volume} {88}},\ \bibinfo {pages} {032913} (\bibinfo
  {year} {2013})}\BibitemShut {NoStop}%
\bibitem [{\citenamefont {Torres-Herrera}\ \emph {et~al.}(2015)\citenamefont
  {Torres-Herrera}, \citenamefont {Kollmar},\ and\ \citenamefont
  {Santos}}]{TorresKollmar2015}%
  \BibitemOpen
  \bibfield  {author} {\bibinfo {author} {\bibfnamefont {E.~J.}\ \bibnamefont
  {Torres-Herrera}}, \bibinfo {author} {\bibfnamefont {D.}~\bibnamefont
  {Kollmar}}, \ and\ \bibinfo {author} {\bibfnamefont {L.~F.}\ \bibnamefont
  {Santos}},\ }\bibfield  {title} {\enquote {\bibinfo {title} {Relaxation and
  thermalization of isolated many-body quantum systems},}\ }\href@noop {}
  {\bibfield  {journal} {\bibinfo  {journal} {Phys. Scr. T}\ }\textbf {\bibinfo
  {volume} {165}},\ \bibinfo {pages} {014018} (\bibinfo {year}
  {2015})}\BibitemShut {NoStop}%
\bibitem [{\citenamefont {Nation}\ and\ \citenamefont
  {Porras}()}]{Nation2019ARXIV}%
  \BibitemOpen
  \bibfield  {author} {\bibinfo {author} {\bibfnamefont {Charlie}\ \bibnamefont
  {Nation}}\ and\ \bibinfo {author} {\bibfnamefont {Diego}\ \bibnamefont
  {Porras}},\ }\href@noop {} {\enquote {\bibinfo {title} {Non-ergodic quantum
  thermalization},}\ }\bibinfo {note} {ArXiv:1908.11773}\BibitemShut {NoStop}%
\bibitem [{\citenamefont {Serbyn}\ \emph {et~al.}(2014)\citenamefont {Serbyn},
  \citenamefont {Papi\ifmmode~\acute{c}\else \'{c}\fi{}},\ and\ \citenamefont
  {Abanin}}]{Serbyn2014}%
  \BibitemOpen
  \bibfield  {author} {\bibinfo {author} {\bibfnamefont {Maksym}\ \bibnamefont
  {Serbyn}}, \bibinfo {author} {\bibfnamefont {Z.}~\bibnamefont
  {Papi\ifmmode~\acute{c}\else \'{c}\fi{}}}, \ and\ \bibinfo {author}
  {\bibfnamefont {D.~A.}\ \bibnamefont {Abanin}},\ }\bibfield  {title}
  {\enquote {\bibinfo {title} {Quantum quenches in the many-body localized
  phase},}\ }\href {\doibase 10.1103/PhysRevB.90.174302} {\bibfield  {journal}
  {\bibinfo  {journal} {Phys. Rev. B}\ }\textbf {\bibinfo {volume} {90}},\
  \bibinfo {pages} {174302} (\bibinfo {year} {2014})}\BibitemShut {NoStop}%
\bibitem [{\citenamefont {Dymarsky}(2019)}]{Dymarsky2019}%
  \BibitemOpen
  \bibfield  {author} {\bibinfo {author} {\bibfnamefont {Anatoly}\ \bibnamefont
  {Dymarsky}},\ }\bibfield  {title} {\enquote {\bibinfo {title} {Mechanism of
  macroscopic equilibration of isolated quantum systems},}\ }\href {\doibase
  10.1103/PhysRevB.99.224302} {\bibfield  {journal} {\bibinfo  {journal} {Phys.
  Rev. B}\ }\textbf {\bibinfo {volume} {99}},\ \bibinfo {pages} {224302}
  (\bibinfo {year} {2019})}\BibitemShut {NoStop}%
\bibitem [{\citenamefont {Dukesz}\ \emph {et~al.}(2009)\citenamefont {Dukesz},
  \citenamefont {Zilbergerts},\ and\ \citenamefont {Santos}}]{Dukesz2009}%
  \BibitemOpen
  \bibfield  {author} {\bibinfo {author} {\bibfnamefont {F.}~\bibnamefont
  {Dukesz}}, \bibinfo {author} {\bibfnamefont {M.}~\bibnamefont {Zilbergerts}},
  \ and\ \bibinfo {author} {\bibfnamefont {L.~F.}\ \bibnamefont {Santos}},\
  }\bibfield  {title} {\enquote {\bibinfo {title} {Interplay between
  interaction and (un)correlated disorder in one-dimensional many-particle
  systems: delocalization and global entanglement},}\ }\href {\doibase
  10.1088/1367-2630/11/4/043026} {\bibfield  {journal} {\bibinfo  {journal}
  {New J. Phys.}\ }\textbf {\bibinfo {volume} {11}},\ \bibinfo {pages} {043026}
  (\bibinfo {year} {2009})}\BibitemShut {NoStop}%
\bibitem [{\citenamefont {Pal}\ and\ \citenamefont {Huse}(2010)}]{Pal2010}%
  \BibitemOpen
  \bibfield  {author} {\bibinfo {author} {\bibfnamefont {A.}~\bibnamefont
  {Pal}}\ and\ \bibinfo {author} {\bibfnamefont {D.~A.}\ \bibnamefont {Huse}},\
  }\bibfield  {title} {\enquote {\bibinfo {title} {Many-body localization phase
  transition},}\ }\href {\doibase 10.1103/PhysRevB.82.174411} {\bibfield
  {journal} {\bibinfo  {journal} {Phys. Rev. B}\ }\textbf {\bibinfo {volume}
  {82}},\ \bibinfo {pages} {174411} (\bibinfo {year} {2010})}\BibitemShut
  {NoStop}%
\bibitem [{\citenamefont {Mac\'e}\ \emph {et~al.}(2019)\citenamefont {Mac\'e},
  \citenamefont {Alet},\ and\ \citenamefont {Laflorencie}}]{Mace2019}%
  \BibitemOpen
  \bibfield  {author} {\bibinfo {author} {\bibfnamefont {Nicolas}\ \bibnamefont
  {Mac\'e}}, \bibinfo {author} {\bibfnamefont {Fabien}\ \bibnamefont {Alet}}, \
  and\ \bibinfo {author} {\bibfnamefont {Nicolas}\ \bibnamefont
  {Laflorencie}},\ }\bibfield  {title} {\enquote {\bibinfo {title}
  {Multifractal scalings across the many-body localization transition},}\
  }\href {\doibase 10.1103/PhysRevLett.123.180601} {\bibfield  {journal}
  {\bibinfo  {journal} {Phys. Rev. Lett.}\ }\textbf {\bibinfo {volume} {123}},\
  \bibinfo {pages} {180601} (\bibinfo {year} {2019})}\BibitemShut {NoStop}%
\bibitem [{\citenamefont {Luitz}\ \emph {et~al.}(2015)\citenamefont {Luitz},
  \citenamefont {Laflorencie},\ and\ \citenamefont {Alet}}]{Luitz2015}%
  \BibitemOpen
  \bibfield  {author} {\bibinfo {author} {\bibfnamefont {D.~J.}\ \bibnamefont
  {Luitz}}, \bibinfo {author} {\bibfnamefont {N.}~\bibnamefont {Laflorencie}},
  \ and\ \bibinfo {author} {\bibfnamefont {F.}~\bibnamefont {Alet}},\
  }\bibfield  {title} {\enquote {\bibinfo {title} {Many-body localization edge
  in the random-field {H}eisenberg chain},}\ }\href {\doibase
  10.1103/PhysRevB.91.081103} {\bibfield  {journal} {\bibinfo  {journal} {Phys.
  Rev. B}\ }\textbf {\bibinfo {volume} {91}},\ \bibinfo {pages} {081103}
  (\bibinfo {year} {2015})}\BibitemShut {NoStop}%
\bibitem [{\citenamefont {Bardarson}\ \emph {et~al.}(2012)\citenamefont
  {Bardarson}, \citenamefont {Pollmann},\ and\ \citenamefont
  {Moore}}]{Bardarson2012}%
  \BibitemOpen
  \bibfield  {author} {\bibinfo {author} {\bibfnamefont {Jens~H.}\ \bibnamefont
  {Bardarson}}, \bibinfo {author} {\bibfnamefont {Frank}\ \bibnamefont
  {Pollmann}}, \ and\ \bibinfo {author} {\bibfnamefont {Joel~E.}\ \bibnamefont
  {Moore}},\ }\bibfield  {title} {\enquote {\bibinfo {title} {Unbounded growth
  of entanglement in models of many-body localization},}\ }\href {\doibase
  10.1103/PhysRevLett.109.017202} {\bibfield  {journal} {\bibinfo  {journal}
  {Phys. Rev. Lett.}\ }\textbf {\bibinfo {volume} {109}},\ \bibinfo {pages}
  {017202} (\bibinfo {year} {2012})}\BibitemShut {NoStop}%
\bibitem [{\citenamefont {Serbyn}\ \emph
  {et~al.}(2013{\natexlab{b}})\citenamefont {Serbyn}, \citenamefont
  {Papi\ifmmode~\acute{c}\else \'{c}\fi{}},\ and\ \citenamefont
  {Abanin}}]{Serbyn_2013_enta}%
  \BibitemOpen
  \bibfield  {author} {\bibinfo {author} {\bibfnamefont {Maksym}\ \bibnamefont
  {Serbyn}}, \bibinfo {author} {\bibfnamefont {Z.}~\bibnamefont
  {Papi\ifmmode~\acute{c}\else \'{c}\fi{}}}, \ and\ \bibinfo {author}
  {\bibfnamefont {Dmitry~A.}\ \bibnamefont {Abanin}},\ }\bibfield  {title}
  {\enquote {\bibinfo {title} {Universal slow growth of entanglement in
  interacting strongly disordered systems},}\ }\href {\doibase
  10.1103/PhysRevLett.110.260601} {\bibfield  {journal} {\bibinfo  {journal}
  {Phys. Rev. Lett.}\ }\textbf {\bibinfo {volume} {110}},\ \bibinfo {pages}
  {260601} (\bibinfo {year} {2013}{\natexlab{b}})}\BibitemShut {NoStop}%
\bibitem [{\citenamefont {Imbrie}(2016)}]{Imbrie2016}%
  \BibitemOpen
  \bibfield  {author} {\bibinfo {author} {\bibfnamefont {John~Z.}\ \bibnamefont
  {Imbrie}},\ }\bibfield  {title} {\enquote {\bibinfo {title} {On many-body
  localization for quantum spin chains},}\ }\href {\doibase
  10.1007/s10955-016-1508-x} {\bibfield  {journal} {\bibinfo  {journal} {J.
  Stat. Phys.}\ }\textbf {\bibinfo {volume} {163}},\ \bibinfo {pages}
  {998--1048} (\bibinfo {year} {2016})}\BibitemShut {NoStop}%
\bibitem [{\citenamefont {Wu}\ \emph {et~al.}(2019)\citenamefont {Wu},
  \citenamefont {Schnell}, \citenamefont {Tomasi}, \citenamefont {Heyl},\ and\
  \citenamefont {Eckardt}}]{Wu_2019}%
  \BibitemOpen
  \bibfield  {author} {\bibinfo {author} {\bibfnamefont {Ling-Na}\ \bibnamefont
  {Wu}}, \bibinfo {author} {\bibfnamefont {Alexander}\ \bibnamefont {Schnell}},
  \bibinfo {author} {\bibfnamefont {Giuseppe~De}\ \bibnamefont {Tomasi}},
  \bibinfo {author} {\bibfnamefont {Markus}\ \bibnamefont {Heyl}}, \ and\
  \bibinfo {author} {\bibfnamefont {Andr{\'{e}}}\ \bibnamefont {Eckardt}},\
  }\bibfield  {title} {\enquote {\bibinfo {title} {Describing many-body
  localized systems in thermal environments},}\ }\href {\doibase
  10.1088/1367-2630/ab25a4} {\bibfield  {journal} {\bibinfo  {journal} {New
  Journal of Physics}\ }\textbf {\bibinfo {volume} {21}},\ \bibinfo {pages}
  {063026} (\bibinfo {year} {2019})}\BibitemShut {NoStop}%
\bibitem [{\citenamefont {De~Tomasi}(2019)}]{DeTomasi2019}%
  \BibitemOpen
  \bibfield  {author} {\bibinfo {author} {\bibfnamefont {Giuseppe}\
  \bibnamefont {De~Tomasi}},\ }\bibfield  {title} {\enquote {\bibinfo {title}
  {Algebraic many-body localization and its implications on information
  propagation},}\ }\href {\doibase 10.1103/PhysRevB.99.054204} {\bibfield
  {journal} {\bibinfo  {journal} {Phys. Rev. B}\ }\textbf {\bibinfo {volume}
  {99}},\ \bibinfo {pages} {054204} (\bibinfo {year} {2019})}\BibitemShut
  {NoStop}%
\bibitem [{\citenamefont {Abanin}\ \emph {et~al.}()\citenamefont {Abanin},
  \citenamefont {Bardarson}, \citenamefont {Tomasi}, \citenamefont
  {Gopalakrishnan}, \citenamefont {Khemani}, \citenamefont {Parameswaran},
  \citenamefont {Pollmann}, \citenamefont {Potter}, \citenamefont {Serbyn},\
  and\ \citenamefont {Vasseur}}]{AbaninARXIV}%
  \BibitemOpen
  \bibfield  {author} {\bibinfo {author} {\bibfnamefont {D.~A.}\ \bibnamefont
  {Abanin}}, \bibinfo {author} {\bibfnamefont {J.~H.}\ \bibnamefont
  {Bardarson}}, \bibinfo {author} {\bibfnamefont {G.~De}\ \bibnamefont
  {Tomasi}}, \bibinfo {author} {\bibfnamefont {S.}~\bibnamefont
  {Gopalakrishnan}}, \bibinfo {author} {\bibfnamefont {V.}~\bibnamefont
  {Khemani}}, \bibinfo {author} {\bibfnamefont {S.~A.}\ \bibnamefont
  {Parameswaran}}, \bibinfo {author} {\bibfnamefont {F.}~\bibnamefont
  {Pollmann}}, \bibinfo {author} {\bibfnamefont {A.~C.}\ \bibnamefont
  {Potter}}, \bibinfo {author} {\bibfnamefont {M.}~\bibnamefont {Serbyn}}, \
  and\ \bibinfo {author} {\bibfnamefont {R.}~\bibnamefont {Vasseur}},\
  }\href@noop {} {\enquote {\bibinfo {title} {Distinguishing localization from
  chaos: challenges in finite-size systems},}\ }\bibinfo {note}
  {ArXiv:1911.04501}\BibitemShut {NoStop}%
\bibitem [{\citenamefont {Khalfin}(1958)}]{Khalfin1958}%
  \BibitemOpen
  \bibfield  {author} {\bibinfo {author} {\bibfnamefont {L.~A.}\ \bibnamefont
  {Khalfin}},\ }\bibfield  {title} {\enquote {\bibinfo {title} {Contribution to
  the decay theory of a quasi-stationary state},}\ }\href@noop {} {\bibfield
  {journal} {\bibinfo  {journal} {Sov. Phys. JETP}\ }\textbf {\bibinfo {volume}
  {6}},\ \bibinfo {pages} {1053} (\bibinfo {year} {1958})}\BibitemShut
  {NoStop}%
\bibitem [{\citenamefont {Fonda}\ \emph {et~al.}(1978)\citenamefont {Fonda},
  \citenamefont {Ghirardi},\ and\ \citenamefont {Rimini}}]{Fonda1978}%
  \BibitemOpen
  \bibfield  {author} {\bibinfo {author} {\bibfnamefont {L.}~\bibnamefont
  {Fonda}}, \bibinfo {author} {\bibfnamefont {G.~C.}\ \bibnamefont {Ghirardi}},
  \ and\ \bibinfo {author} {\bibfnamefont {A.}~\bibnamefont {Rimini}},\
  }\bibfield  {title} {\enquote {\bibinfo {title} {Decay theory of unstable
  quantum systems},}\ }\href@noop {} {\bibfield  {journal} {\bibinfo  {journal}
  {Rep. Prog. Phys.}\ }\textbf {\bibinfo {volume} {41}},\ \bibinfo {pages}
  {587} (\bibinfo {year} {1978})}\BibitemShut {NoStop}%
\bibitem [{\citenamefont {Bhattacharyya}(1983)}]{Bhattacharyya1983}%
  \BibitemOpen
  \bibfield  {author} {\bibinfo {author} {\bibfnamefont {K.}~\bibnamefont
  {Bhattacharyya}},\ }\bibfield  {title} {\enquote {\bibinfo {title} {{Q}uantum
  decay and the {M}andelstam-{T}amm-energy inequality},}\ }\href@noop {}
  {\bibfield  {journal} {\bibinfo  {journal} {J. Phys. A}\ }\textbf {\bibinfo
  {volume} {16}},\ \bibinfo {pages} {2993} (\bibinfo {year}
  {1983})}\BibitemShut {NoStop}%
\bibitem [{\citenamefont {Ketzmerick}\ \emph {et~al.}(1992)\citenamefont
  {Ketzmerick}, \citenamefont {Petschel},\ and\ \citenamefont
  {Geisel}}]{Ketzmerick1992}%
  \BibitemOpen
  \bibfield  {author} {\bibinfo {author} {\bibfnamefont {R.}~\bibnamefont
  {Ketzmerick}}, \bibinfo {author} {\bibfnamefont {G.}~\bibnamefont
  {Petschel}}, \ and\ \bibinfo {author} {\bibfnamefont {T.}~\bibnamefont
  {Geisel}},\ }\bibfield  {title} {\enquote {\bibinfo {title} {Slow decay of
  temporal correlations in quantum systems with {C}antor spectra},}\ }\href
  {\doibase 10.1103/PhysRevLett.69.695} {\bibfield  {journal} {\bibinfo
  {journal} {Phys. Rev. Lett.}\ }\textbf {\bibinfo {volume} {69}},\ \bibinfo
  {pages} {695--698} (\bibinfo {year} {1992})}\BibitemShut {NoStop}%
\bibitem [{\citenamefont {Muga}\ \emph {et~al.}(2009)\citenamefont {Muga},
  \citenamefont {Ruschhaupt},\ and\ \citenamefont {del Campo}}]{MugaBook}%
  \BibitemOpen
  \bibfield  {author} {\bibinfo {author} {\bibfnamefont {J.~G.}\ \bibnamefont
  {Muga}}, \bibinfo {author} {\bibfnamefont {A.}~\bibnamefont {Ruschhaupt}}, \
  and\ \bibinfo {author} {\bibfnamefont {A.}~\bibnamefont {del Campo}},\
  }\href@noop {} {\emph {\bibinfo {title} {Time in Quantum Mechanics, vol.
  2}}}\ (\bibinfo  {publisher} {Springer},\ \bibinfo {address} {London},\
  \bibinfo {year} {2009})\BibitemShut {NoStop}%
\bibitem [{\citenamefont {Torres-Herrera}\ and\ \citenamefont
  {Santos}(2014{\natexlab{a}})}]{Torres2014PRA}%
  \BibitemOpen
  \bibfield  {author} {\bibinfo {author} {\bibfnamefont {E.~J.}\ \bibnamefont
  {Torres-Herrera}}\ and\ \bibinfo {author} {\bibfnamefont {Lea~F.}\
  \bibnamefont {Santos}},\ }\bibfield  {title} {\enquote {\bibinfo {title}
  {Quench dynamics of isolated many-body quantum systems},}\ }\href {\doibase
  10.1103/PhysRevA.89.043620} {\bibfield  {journal} {\bibinfo  {journal} {Phys.
  Rev. A}\ }\textbf {\bibinfo {volume} {89}},\ \bibinfo {pages} {043620}
  (\bibinfo {year} {2014}{\natexlab{a}})}\BibitemShut {NoStop}%
\bibitem [{\citenamefont {Torres-Herrera}\ \emph {et~al.}(2014)\citenamefont
  {Torres-Herrera}, \citenamefont {Vyas},\ and\ \citenamefont
  {Santos}}]{Torres2014NJP}%
  \BibitemOpen
  \bibfield  {author} {\bibinfo {author} {\bibfnamefont {E.~J.}\ \bibnamefont
  {Torres-Herrera}}, \bibinfo {author} {\bibfnamefont {M}~\bibnamefont {Vyas}},
  \ and\ \bibinfo {author} {\bibfnamefont {Lea~F.}\ \bibnamefont {Santos}},\
  }\bibfield  {title} {\enquote {\bibinfo {title} {General features of the
  relaxation dynamics of interacting quantum systems},}\ }\href@noop {}
  {\bibfield  {journal} {\bibinfo  {journal} {New J. Phys.}\ }\textbf {\bibinfo
  {volume} {16}},\ \bibinfo {pages} {063010} (\bibinfo {year}
  {2014})}\BibitemShut {NoStop}%
\bibitem [{\citenamefont {Torres-Herrera}\ and\ \citenamefont
  {Santos}(2014{\natexlab{b}})}]{Torres2014PRE}%
  \BibitemOpen
  \bibfield  {author} {\bibinfo {author} {\bibfnamefont {E.~J.}\ \bibnamefont
  {Torres-Herrera}}\ and\ \bibinfo {author} {\bibfnamefont {Lea~F.}\
  \bibnamefont {Santos}},\ }\bibfield  {title} {\enquote {\bibinfo {title}
  {Local quenches with global effects in interacting quantum systems},}\ }\href
  {\doibase 10.1103/PhysRevE.89.062110} {\bibfield  {journal} {\bibinfo
  {journal} {Phys. Rev. E}\ }\textbf {\bibinfo {volume} {89}},\ \bibinfo
  {pages} {062110} (\bibinfo {year} {2014}{\natexlab{b}})}\BibitemShut
  {NoStop}%
\bibitem [{\citenamefont {Torres-Herrera}\ and\ \citenamefont
  {Santos}(2014{\natexlab{c}})}]{Torres2014PRAb}%
  \BibitemOpen
  \bibfield  {author} {\bibinfo {author} {\bibfnamefont {E.~J.}\ \bibnamefont
  {Torres-Herrera}}\ and\ \bibinfo {author} {\bibfnamefont {Lea~F.}\
  \bibnamefont {Santos}},\ }\bibfield  {title} {\enquote {\bibinfo {title}
  {Nonexponential fidelity decay in isolated interacting quantum systems},}\
  }\href {\doibase 10.1103/PhysRevA.90.033623} {\bibfield  {journal} {\bibinfo
  {journal} {Phys. Rev. A}\ }\textbf {\bibinfo {volume} {90}},\ \bibinfo
  {pages} {033623} (\bibinfo {year} {2014}{\natexlab{c}})}\BibitemShut
  {NoStop}%
\bibitem [{\citenamefont {Mazza}\ \emph {et~al.}(2016)\citenamefont {Mazza},
  \citenamefont {St{\'{e}}phan}, \citenamefont {Canovi}, \citenamefont {Alba},
  \citenamefont {Brockmann},\ and\ \citenamefont {Haque}}]{Mazza2016}%
  \BibitemOpen
  \bibfield  {author} {\bibinfo {author} {\bibfnamefont {Paolo~P}\ \bibnamefont
  {Mazza}}, \bibinfo {author} {\bibfnamefont {Jean-Marie}\ \bibnamefont
  {St{\'{e}}phan}}, \bibinfo {author} {\bibfnamefont {Elena}\ \bibnamefont
  {Canovi}}, \bibinfo {author} {\bibfnamefont {Vincenzo}\ \bibnamefont {Alba}},
  \bibinfo {author} {\bibfnamefont {Michael}\ \bibnamefont {Brockmann}}, \ and\
  \bibinfo {author} {\bibfnamefont {Masudul}\ \bibnamefont {Haque}},\
  }\bibfield  {title} {\enquote {\bibinfo {title} {Overlap distributions for
  quantum quenches in the anisotropic {H}eisenberg chain},}\ }\href {\doibase
  10.1088/1742-5468/2016/01/013104} {\bibfield  {journal} {\bibinfo  {journal}
  {J. Stat. Mech.}\ }\textbf {\bibinfo {volume} {2016}},\ \bibinfo {pages}
  {013104} (\bibinfo {year} {2016})}\BibitemShut {NoStop}%
\bibitem [{\citenamefont {Torres-Herrera}\ \emph {et~al.}(2018)\citenamefont
  {Torres-Herrera}, \citenamefont {Garc\'{\i}a-Garc\'{\i}a},\ and\
  \citenamefont {Santos}}]{Torres2018}%
  \BibitemOpen
  \bibfield  {author} {\bibinfo {author} {\bibfnamefont {E.~J.}\ \bibnamefont
  {Torres-Herrera}}, \bibinfo {author} {\bibfnamefont {Antonio~M.}\
  \bibnamefont {Garc\'{\i}a-Garc\'{\i}a}}, \ and\ \bibinfo {author}
  {\bibfnamefont {Lea~F.}\ \bibnamefont {Santos}},\ }\bibfield  {title}
  {\enquote {\bibinfo {title} {Generic dynamical features of quenched
  interacting quantum systems: Survival probability, density imbalance, and
  out-of-time-ordered correlator},}\ }\href {\doibase
  10.1103/PhysRevB.97.060303} {\bibfield  {journal} {\bibinfo  {journal} {Phys.
  Rev. B}\ }\textbf {\bibinfo {volume} {97}},\ \bibinfo {pages} {060303}
  (\bibinfo {year} {2018})}\BibitemShut {NoStop}%
\bibitem [{\citenamefont {Volya}\ and\ \citenamefont
  {Zelevinsky}()}]{VolyaARXIV}%
  \BibitemOpen
  \bibfield  {author} {\bibinfo {author} {\bibfnamefont {A.}~\bibnamefont
  {Volya}}\ and\ \bibinfo {author} {\bibfnamefont {V.}~\bibnamefont
  {Zelevinsky}},\ }\href@noop {} {\enquote {\bibinfo {title} {Time-dependent
  relaxation of observables in complex quantum systems},}\ }\bibinfo {note}
  {ArXiv:1905.11918}\BibitemShut {NoStop}%
\bibitem [{\citenamefont {Bera}\ \emph {et~al.}(2018)\citenamefont {Bera},
  \citenamefont {De~Tomasi}, \citenamefont {Khaymovich},\ and\ \citenamefont
  {Scardicchio}}]{Bera2018}%
  \BibitemOpen
  \bibfield  {author} {\bibinfo {author} {\bibfnamefont {Soumya}\ \bibnamefont
  {Bera}}, \bibinfo {author} {\bibfnamefont {Giuseppe}\ \bibnamefont
  {De~Tomasi}}, \bibinfo {author} {\bibfnamefont {Ivan~M.}\ \bibnamefont
  {Khaymovich}}, \ and\ \bibinfo {author} {\bibfnamefont {Antonello}\
  \bibnamefont {Scardicchio}},\ }\bibfield  {title} {\enquote {\bibinfo {title}
  {Return probability for the {A}nderson model on the random regular graph},}\
  }\href {\doibase 10.1103/PhysRevB.98.134205} {\bibfield  {journal} {\bibinfo
  {journal} {Phys. Rev. B}\ }\textbf {\bibinfo {volume} {98}},\ \bibinfo
  {pages} {134205} (\bibinfo {year} {2018})}\BibitemShut {NoStop}%
\bibitem [{\citenamefont {Lerma-Hern{\'{a}}ndez}\ \emph
  {et~al.}(2018)\citenamefont {Lerma-Hern{\'{a}}ndez}, \citenamefont
  {Ch{\'{a}}vez-Carlos}, \citenamefont {Bastarrachea-Magnani}, \citenamefont
  {Santos},\ and\ \citenamefont {Hirsch}}]{Lerma2018}%
  \BibitemOpen
  \bibfield  {author} {\bibinfo {author} {\bibfnamefont {Sergio}\ \bibnamefont
  {Lerma-Hern{\'{a}}ndez}}, \bibinfo {author} {\bibfnamefont {Jorge}\
  \bibnamefont {Ch{\'{a}}vez-Carlos}}, \bibinfo {author} {\bibfnamefont
  {Miguel~A}\ \bibnamefont {Bastarrachea-Magnani}}, \bibinfo {author}
  {\bibfnamefont {Lea~F}\ \bibnamefont {Santos}}, \ and\ \bibinfo {author}
  {\bibfnamefont {Jorge~G}\ \bibnamefont {Hirsch}},\ }\bibfield  {title}
  {\enquote {\bibinfo {title} {Analytical description of the survival
  probability of coherent states in regular regimes},}\ }\href {\doibase
  10.1088/1751-8121/aae2c3} {\bibfield  {journal} {\bibinfo  {journal} {J.
  Phys. A}\ }\textbf {\bibinfo {volume} {51}},\ \bibinfo {pages} {475302}
  (\bibinfo {year} {2018})}\BibitemShut {NoStop}%
\bibitem [{\citenamefont {Reimann}(2016)}]{Reimann2016}%
  \BibitemOpen
  \bibfield  {author} {\bibinfo {author} {\bibfnamefont {Peter}\ \bibnamefont
  {Reimann}},\ }\bibfield  {title} {\enquote {\bibinfo {title} {Typical fast
  thermalization processes in closed many-body systems},}\ }\href {\doibase
  http://dx.doi.org/10.1038/ncomms10821} {\bibfield  {journal} {\bibinfo
  {journal} {Nat. Comm.}\ }\textbf {\bibinfo {volume} {7}},\ \bibinfo {pages}
  {10821} (\bibinfo {year} {2016})}\BibitemShut {NoStop}%
\bibitem [{\citenamefont {Reimann}(2019)}]{Reimann2019}%
  \BibitemOpen
  \bibfield  {author} {\bibinfo {author} {\bibfnamefont {Peter}\ \bibnamefont
  {Reimann}},\ }\bibfield  {title} {\enquote {\bibinfo {title} {Transportless
  equilibration in isolated many-body quantum systems},}\ }\href {\doibase
  10.1088/1367-2630/ab1a63} {\bibfield  {journal} {\bibinfo  {journal} {New J.
  Phys.}\ }\textbf {\bibinfo {volume} {21}},\ \bibinfo {pages} {053014}
  (\bibinfo {year} {2019})}\BibitemShut {NoStop}%
\bibitem [{\citenamefont {Tomasi}\ \emph {et~al.}(2019)\citenamefont {Tomasi},
  \citenamefont {Amini}, \citenamefont {Bera}, \citenamefont {Khaymovich},\
  and\ \citenamefont {Kravtsov}}]{DeTomasi_2019_survival}%
  \BibitemOpen
  \bibfield  {author} {\bibinfo {author} {\bibfnamefont {G.~De}\ \bibnamefont
  {Tomasi}}, \bibinfo {author} {\bibfnamefont {M.}~\bibnamefont {Amini}},
  \bibinfo {author} {\bibfnamefont {S.}~\bibnamefont {Bera}}, \bibinfo {author}
  {\bibfnamefont {I.~M.}\ \bibnamefont {Khaymovich}}, \ and\ \bibinfo {author}
  {\bibfnamefont {V.~E.}\ \bibnamefont {Kravtsov}},\ }\bibfield  {title}
  {\enquote {\bibinfo {title} {{Survival probability in Generalized
  Rosenzweig-Porter random matrix ensemble}},}\ }\href {\doibase
  10.21468/SciPostPhys.6.1.014} {\bibfield  {journal} {\bibinfo  {journal}
  {SciPost Phys.}\ }\textbf {\bibinfo {volume} {6}},\ \bibinfo {pages} {14}
  (\bibinfo {year} {2019})}\BibitemShut {NoStop}%
\bibitem [{\citenamefont {Singh}\ \emph {et~al.}()\citenamefont {Singh},
  \citenamefont {Fujiwara}, \citenamefont {Geiger}, \citenamefont {Simmons},
  \citenamefont {Lipatov}, \citenamefont {Cao}, \citenamefont {Dotti},
  \citenamefont {Rajagopal}, \citenamefont {Senaratne}, \citenamefont
  {Shimasaki}, \citenamefont {Heyl}, \citenamefont {Eckardt},\ and\
  \citenamefont {Weld}}]{SinghARXIV}%
  \BibitemOpen
  \bibfield  {author} {\bibinfo {author} {\bibfnamefont {Kevin}\ \bibnamefont
  {Singh}}, \bibinfo {author} {\bibfnamefont {Cora~J.}\ \bibnamefont
  {Fujiwara}}, \bibinfo {author} {\bibfnamefont {Zachary~A.}\ \bibnamefont
  {Geiger}}, \bibinfo {author} {\bibfnamefont {Ethan~Q.}\ \bibnamefont
  {Simmons}}, \bibinfo {author} {\bibfnamefont {Mikhail}\ \bibnamefont
  {Lipatov}}, \bibinfo {author} {\bibfnamefont {Alec}\ \bibnamefont {Cao}},
  \bibinfo {author} {\bibfnamefont {Peter}\ \bibnamefont {Dotti}}, \bibinfo
  {author} {\bibfnamefont {Shankari~V.}\ \bibnamefont {Rajagopal}}, \bibinfo
  {author} {\bibfnamefont {Ruwan}\ \bibnamefont {Senaratne}}, \bibinfo {author}
  {\bibfnamefont {Toshihiko}\ \bibnamefont {Shimasaki}}, \bibinfo {author}
  {\bibfnamefont {Markus}\ \bibnamefont {Heyl}}, \bibinfo {author}
  {\bibfnamefont {Andr\'e}\ \bibnamefont {Eckardt}}, \ and\ \bibinfo {author}
  {\bibfnamefont {David~M.}\ \bibnamefont {Weld}},\ }\href@noop {} {\enquote
  {\bibinfo {title} {Quantifying and controlling prethermal nonergodicity in
  interacting {F}loquet matter},}\ }\bibinfo {note}
  {ArXiv:1809.05554}\BibitemShut {NoStop}%
\bibitem [{\citenamefont {Borgonovi}\ \emph
  {et~al.}(2019{\natexlab{a}})\citenamefont {Borgonovi}, \citenamefont
  {Izrailev},\ and\ \citenamefont {Santos}}]{Borgonovi2019}%
  \BibitemOpen
  \bibfield  {author} {\bibinfo {author} {\bibfnamefont {Fausto}\ \bibnamefont
  {Borgonovi}}, \bibinfo {author} {\bibfnamefont {Felix~M.}\ \bibnamefont
  {Izrailev}}, \ and\ \bibinfo {author} {\bibfnamefont {Lea~F.}\ \bibnamefont
  {Santos}},\ }\bibfield  {title} {\enquote {\bibinfo {title} {Exponentially
  fast dynamics of chaotic many-body systems},}\ }\href {\doibase
  10.1103/PhysRevE.99.010101} {\bibfield  {journal} {\bibinfo  {journal} {Phys.
  Rev. E}\ }\textbf {\bibinfo {volume} {99}},\ \bibinfo {pages} {010101}
  (\bibinfo {year} {2019}{\natexlab{a}})}\BibitemShut {NoStop}%
\bibitem [{\citenamefont {Borgonovi}\ \emph
  {et~al.}(2019{\natexlab{b}})\citenamefont {Borgonovi}, \citenamefont
  {Izrailev},\ and\ \citenamefont {Santos}}]{Borgonovi2019b}%
  \BibitemOpen
  \bibfield  {author} {\bibinfo {author} {\bibfnamefont {Fausto}\ \bibnamefont
  {Borgonovi}}, \bibinfo {author} {\bibfnamefont {Felix~M.}\ \bibnamefont
  {Izrailev}}, \ and\ \bibinfo {author} {\bibfnamefont {Lea~F.}\ \bibnamefont
  {Santos}},\ }\bibfield  {title} {\enquote {\bibinfo {title} {Timescales in
  the quench dynamics of many-body quantum systems: Participation ratio versus
  out-of-time ordered correlator},}\ }\href {\doibase
  10.1103/PhysRevE.99.052143} {\bibfield  {journal} {\bibinfo  {journal} {Phys.
  Rev. E}\ }\textbf {\bibinfo {volume} {99}},\ \bibinfo {pages} {052143}
  (\bibinfo {year} {2019}{\natexlab{b}})}\BibitemShut {NoStop}%
\bibitem [{\citenamefont {De~Luca}\ \emph {et~al.}(2014)\citenamefont
  {De~Luca}, \citenamefont {Altshuler}, \citenamefont {Kravtsov},\ and\
  \citenamefont {Scardicchio}}]{Luca2014}%
  \BibitemOpen
  \bibfield  {author} {\bibinfo {author} {\bibfnamefont {A.}~\bibnamefont
  {De~Luca}}, \bibinfo {author} {\bibfnamefont {B.~L.}\ \bibnamefont
  {Altshuler}}, \bibinfo {author} {\bibfnamefont {V.~E.}\ \bibnamefont
  {Kravtsov}}, \ and\ \bibinfo {author} {\bibfnamefont {A.}~\bibnamefont
  {Scardicchio}},\ }\bibfield  {title} {\enquote {\bibinfo {title} {Anderson
  localization on the {B}ethe lattice: Nonergodicity of extended states},}\
  }\href {\doibase 10.1103/PhysRevLett.113.046806} {\bibfield  {journal}
  {\bibinfo  {journal} {Phys. Rev. Lett.}\ }\textbf {\bibinfo {volume} {113}},\
  \bibinfo {pages} {046806} (\bibinfo {year} {2014})}\BibitemShut {NoStop}%
\bibitem [{\citenamefont {Torres-Herrera}\ and\ \citenamefont
  {Santos}(2015)}]{Torres2015}%
  \BibitemOpen
  \bibfield  {author} {\bibinfo {author} {\bibfnamefont {E.~J.}\ \bibnamefont
  {Torres-Herrera}}\ and\ \bibinfo {author} {\bibfnamefont {Lea~F.}\
  \bibnamefont {Santos}},\ }\bibfield  {title} {\enquote {\bibinfo {title}
  {Dynamics at the many-body localization transition},}\ }\href {\doibase
  10.1103/PhysRevB.92.014208} {\bibfield  {journal} {\bibinfo  {journal} {Phys.
  Rev. B}\ }\textbf {\bibinfo {volume} {92}},\ \bibinfo {pages} {014208}
  (\bibinfo {year} {2015})}\BibitemShut {NoStop}%
\bibitem [{\citenamefont {Torres-Herrera}\ and\ \citenamefont
  {Santos}(2017{\natexlab{a}})}]{Torres2017}%
  \BibitemOpen
  \bibfield  {author} {\bibinfo {author} {\bibfnamefont {E.~J.}\ \bibnamefont
  {Torres-Herrera}}\ and\ \bibinfo {author} {\bibfnamefont {Lea~F.}\
  \bibnamefont {Santos}},\ }\bibfield  {title} {\enquote {\bibinfo {title}
  {Extended nonergodic states in disordered many-body quantum systems},}\
  }\href {\doibase 10.1002/andp.201600284} {\bibfield  {journal} {\bibinfo
  {journal} {Ann. Phys. (Berlin)}\ }\textbf {\bibinfo {volume} {529}},\
  \bibinfo {pages} {1600284} (\bibinfo {year}
  {2017}{\natexlab{a}})}\BibitemShut {NoStop}%
\bibitem [{\citenamefont {Pino}\ \emph {et~al.}(2017)\citenamefont {Pino},
  \citenamefont {Kravtsov}, \citenamefont {Altshuler},\ and\ \citenamefont
  {Ioffe}}]{Pino2017}%
  \BibitemOpen
  \bibfield  {author} {\bibinfo {author} {\bibfnamefont {M.}~\bibnamefont
  {Pino}}, \bibinfo {author} {\bibfnamefont {V.~E.}\ \bibnamefont {Kravtsov}},
  \bibinfo {author} {\bibfnamefont {B.~L.}\ \bibnamefont {Altshuler}}, \ and\
  \bibinfo {author} {\bibfnamefont {L.~B.}\ \bibnamefont {Ioffe}},\ }\bibfield
  {title} {\enquote {\bibinfo {title} {Multifractal metal in a disordered
  josephson junctions array},}\ }\href {\doibase 10.1103/PhysRevB.96.214205}
  {\bibfield  {journal} {\bibinfo  {journal} {Phys. Rev. B}\ }\textbf {\bibinfo
  {volume} {96}},\ \bibinfo {pages} {214205} (\bibinfo {year}
  {2017})}\BibitemShut {NoStop}%
\bibitem [{\citenamefont {T\'avora}\ \emph {et~al.}(2016)\citenamefont
  {T\'avora}, \citenamefont {Torres-Herrera},\ and\ \citenamefont
  {Santos}}]{Tavora2016}%
  \BibitemOpen
  \bibfield  {author} {\bibinfo {author} {\bibfnamefont {Marco}\ \bibnamefont
  {T\'avora}}, \bibinfo {author} {\bibfnamefont {E.~J.}\ \bibnamefont
  {Torres-Herrera}}, \ and\ \bibinfo {author} {\bibfnamefont {Lea~F.}\
  \bibnamefont {Santos}},\ }\bibfield  {title} {\enquote {\bibinfo {title}
  {Inevitable power-law behavior of isolated many-body quantum systems and how
  it anticipates thermalization},}\ }\href {\doibase
  10.1103/PhysRevA.94.041603} {\bibfield  {journal} {\bibinfo  {journal} {Phys.
  Rev. A}\ }\textbf {\bibinfo {volume} {94}},\ \bibinfo {pages} {041603}
  (\bibinfo {year} {2016})}\BibitemShut {NoStop}%
\bibitem [{\citenamefont {T\'avora}\ \emph {et~al.}(2017)\citenamefont
  {T\'avora}, \citenamefont {Torres-Herrera},\ and\ \citenamefont
  {Santos}}]{Tavora2017}%
  \BibitemOpen
  \bibfield  {author} {\bibinfo {author} {\bibfnamefont {Marco}\ \bibnamefont
  {T\'avora}}, \bibinfo {author} {\bibfnamefont {E.~J.}\ \bibnamefont
  {Torres-Herrera}}, \ and\ \bibinfo {author} {\bibfnamefont {Lea~F.}\
  \bibnamefont {Santos}},\ }\bibfield  {title} {\enquote {\bibinfo {title}
  {Power-law decay exponents: A dynamical criterion for predicting
  thermalization},}\ }\href {\doibase 10.1103/PhysRevA.95.013604} {\bibfield
  {journal} {\bibinfo  {journal} {Phys. Rev. A}\ }\textbf {\bibinfo {volume}
  {95}},\ \bibinfo {pages} {013604} (\bibinfo {year} {2017})}\BibitemShut
  {NoStop}%
\bibitem [{\citenamefont {Leviandier}\ \emph {et~al.}(1986)\citenamefont
  {Leviandier}, \citenamefont {Lombardi}, \citenamefont {Jost},\ and\
  \citenamefont {Pique}}]{Leviandier1986}%
  \BibitemOpen
  \bibfield  {author} {\bibinfo {author} {\bibfnamefont {Luc}\ \bibnamefont
  {Leviandier}}, \bibinfo {author} {\bibfnamefont {Maurice}\ \bibnamefont
  {Lombardi}}, \bibinfo {author} {\bibfnamefont {R\'emi}\ \bibnamefont {Jost}},
  \ and\ \bibinfo {author} {\bibfnamefont {Jean~Paul}\ \bibnamefont {Pique}},\
  }\bibfield  {title} {\enquote {\bibinfo {title} {Fourier transform: A tool to
  measure statistical level properties in very complex spectra},}\ }\href
  {\doibase 10.1103/PhysRevLett.56.2449} {\bibfield  {journal} {\bibinfo
  {journal} {Phys. Rev. Lett.}\ }\textbf {\bibinfo {volume} {56}},\ \bibinfo
  {pages} {2449--2452} (\bibinfo {year} {1986})}\BibitemShut {NoStop}%
\bibitem [{\citenamefont {Schiulaz}\ \emph {et~al.}(2019)\citenamefont
  {Schiulaz}, \citenamefont {Torres-Herrera},\ and\ \citenamefont
  {Santos}}]{Schiulaz2019}%
  \BibitemOpen
  \bibfield  {author} {\bibinfo {author} {\bibfnamefont {Mauro}\ \bibnamefont
  {Schiulaz}}, \bibinfo {author} {\bibfnamefont {E.~Jonathan}\ \bibnamefont
  {Torres-Herrera}}, \ and\ \bibinfo {author} {\bibfnamefont {Lea~F.}\
  \bibnamefont {Santos}},\ }\bibfield  {title} {\enquote {\bibinfo {title}
  {Thouless and relaxation time scales in many-body quantum systems},}\ }\href
  {\doibase 10.1103/PhysRevB.99.174313} {\bibfield  {journal} {\bibinfo
  {journal} {Phys. Rev. B}\ }\textbf {\bibinfo {volume} {99}},\ \bibinfo
  {pages} {174313} (\bibinfo {year} {2019})}\BibitemShut {NoStop}%
\bibitem [{foo()}]{footSierant}%
  \BibitemOpen
  \href@noop {} {}\bibinfo {note} {We call attention about some very
  interesting discussions about the Thouless time in
  Ref.~\cite{Sierant2019ARXIV}}\BibitemShut {NoStop}%
\bibitem [{\citenamefont {Torres-Herrera}\ and\ \citenamefont
  {Santos}(2017{\natexlab{b}})}]{Torres2017Philo}%
  \BibitemOpen
  \bibfield  {author} {\bibinfo {author} {\bibfnamefont {E.~J.}\ \bibnamefont
  {Torres-Herrera}}\ and\ \bibinfo {author} {\bibfnamefont {Lea~F.}\
  \bibnamefont {Santos}},\ }\bibfield  {title} {\enquote {\bibinfo {title}
  {Dynamical manifestations of quantum chaos: correlation hole and bulge},}\
  }\href {\doibase 10.1098/rsta.2016.0434} {\bibfield  {journal} {\bibinfo
  {journal} {Philos. Trans. Royal Soc. A}\ }\textbf {\bibinfo {volume} {375}},\
  \bibinfo {pages} {20160434} (\bibinfo {year}
  {2017}{\natexlab{b}})}\BibitemShut {NoStop}%
\bibitem [{\citenamefont {Torres-Herrera}\ \emph {et~al.}()\citenamefont
  {Torres-Herrera}, \citenamefont {Vallejo-Fabila}, \citenamefont
  {Mart\'{\i}nez-Mendoza},\ and\ \citenamefont {Santos}}]{TorresARXIV}%
  \BibitemOpen
  \bibfield  {author} {\bibinfo {author} {\bibfnamefont {E.~Jonathan}\
  \bibnamefont {Torres-Herrera}}, \bibinfo {author} {\bibfnamefont
  {Isa\'{\i}as}\ \bibnamefont {Vallejo-Fabila}}, \bibinfo {author}
  {\bibfnamefont {Andrei~J.}\ \bibnamefont {Mart\'{\i}nez-Mendoza}}, \ and\
  \bibinfo {author} {\bibfnamefont {Lea~F.}\ \bibnamefont {Santos}},\
  }\href@noop {} {\enquote {\bibinfo {title} {Self-averaging in many-body
  quantum systems out of equilibrium: Time dependence of distributions},}\
  }\bibinfo {note} {ArXiv:2005.14188}\BibitemShut {NoStop}%
\bibitem [{\citenamefont {Sierant}\ \emph {et~al.}()\citenamefont {Sierant},
  \citenamefont {Delande},\ and\ \citenamefont
  {Zakrzewski}}]{Sierant2019ARXIV}%
  \BibitemOpen
  \bibfield  {author} {\bibinfo {author} {\bibfnamefont {Piotr}\ \bibnamefont
  {Sierant}}, \bibinfo {author} {\bibfnamefont {Dominique}\ \bibnamefont
  {Delande}}, \ and\ \bibinfo {author} {\bibfnamefont {Jakub}\ \bibnamefont
  {Zakrzewski}},\ }\href@noop {} {\enquote {\bibinfo {title} {Thouless time
  analysis of {A}nderson and many-body localization transitions},}\ }\bibinfo
  {note} {ArXiv:1911.06221}\BibitemShut {NoStop}%
\end{thebibliography}

%

\end{document}